\documentclass[aps, prl,twocolumn, superscriptaddress, preprintnumbers, floatfix, longbibliography, 10pt]{revtex4-2}

\newcommand\varpm{\mathbin{\vcenter{\hbox{%
  \oalign{\hfil$\scriptstyle+$\hfil\cr
          \noalign{\kern-.3ex}
          $\scriptscriptstyle({-})$\cr}%
      }}}}
\usepackage{graphicx}
\usepackage{dcolumn}
\usepackage{bm}
\usepackage{dsfont}
\usepackage{physics}
\usepackage{xcolor}
\usepackage[caption=false]{subfig}

\usepackage{comment}
\usepackage[colorlinks, allcolors=blue]{hyperref}

\newcommand{\RE}[1]{\,\mbox{Re}\left\{{#1}\right\}}
\newcommand{\IM}[1]{\,\mbox{Im}\left\{{#1}\right\}}
\newcommand{\df}[1]{\,\delta{\left(#1\right)}}

\definecolor{Red}{rgb}{1,0,0}
\definecolor{Blue}{rgb}{0.5,0,0.9}
\definecolor{Green}{rgb}{0,0.5,0}
\definecolor{Magenta}{rgb}{1,0,0.56}
\definecolor{Orange}{rgb}{1,0.64,0}

\begin{document}


\title{Minimal Models and Transport Properties of Unconventional $p$-Wave Magnets}
\author{Bj{\o}rnulf Brekke}
\thanks{These authors contributed equally to this work}
\author{Pavlo Sukhachov}
\thanks{These authors contributed equally to this work}
\affiliation{Center for Quantum Spintronics, Department of Physics, Norwegian \\ University of Science and Technology, NO-7491 Trondheim, Norway}
\author{Hans Gl{\o}ckner Giil}
\affiliation{Center for Quantum Spintronics, Department of Physics, Norwegian \\ University of Science and Technology, NO-7491 Trondheim, Norway}
\author{Arne Brataas}
\affiliation{Center for Quantum Spintronics, Department of Physics, Norwegian \\ University of Science and Technology, NO-7491 Trondheim, Norway}
\author{Jacob Linder}
\affiliation{Center for Quantum Spintronics, Department of Physics, Norwegian \\ University of Science and Technology, NO-7491 Trondheim, Norway}

\date{December 10, 2024}

\begin{abstract}
New unconventional compensated magnets with a $p$-wave spin polarization protected by a composite time-reversal translation symmetry have been proposed in the wake of altermagnets. To facilitate the experimental discovery and applications of these unconventional magnets, we construct an effective analytical model. The effective model is based on a minimal tight-binding model for unconventional $p$-wave magnets that clarifies the relation to other magnets with $p$-wave spin-polarized bands. One of the most prominent advantages of our analytical model is the possibility to employ various analytical approaches while capturing essential features of $p$-wave magnets. We illustrate the effective model by evaluating the tunneling conductance in junctions with $p$-wave magnets, revealing a large magnetoresistance, spin filtering, and anisotropic bulk spin conductivity beyond linear response despite the absence of a net magnetization. These results show that unconventional $p$-wave magnets offer several useful functionalities, broadening the material selection for spintronics devices.
\end{abstract}

\maketitle

\textit{Introduction.---}
Recently, new classes of compensated magnetic materials have attracted significant attention. Among them, one distinguishes magnetic materials with collinear magnetic order from noncollinear and noncoplanar magnets. The former are known as altermagnets~\cite{noda2016momentum, vsmejkal2020crystal, hayami2019momentum, AhnPRB2019, YuanPRB2020, YuanPRM2021, SmejkalPRX2022, SmejkalSinovaPRX2022, mazin_prx_22, bai_arxiv_24} that despite being compensated, feature fully spin-polarized itinerant electrons with a large nonrelativistic even-parity spin splitting, rendering the spin degree of freedom accessible for spin transport~\cite{naka2019spin, GonzalezPRL2021Splitter,shao2021spin, SmejkalPRX22Hellenes, ShutaroPRL2022, Das-Soori:2023}, spin caloritronics~\cite{ZhouPRL2024, CuiPRB2023, liu2024inverse, hoyer2024spontaneous, badura2024observation}, and superconductivity \cite{OuassouPRL2023, beenakker_prb_23, papaj_prb_23, SunPRB2023, Zhu2023PRB, BrekkePRB2023, chakraborty_arxiv_23, ghorashi_arxiv_23, GiilPRB2024, das_arxiv_24, maeland_prb_24, giil2024quasiclassical, zhang2024finite, sukhachov2024thermoelectric, chourasia_arxiv_24}. Many material candidates, such as Ru$_2$O~\footnote{The magnetic order in RuO$_2$ is debated~\cite{Hiraishi-Hiroi-NonmagneticGroundState-2024, Kessler-Moser-Ru02:2024}.}, MnF$_2$, MnTe, and CrSb have been suggested, see also Ref.~\cite{SmejkalPRX2022}. Experimental verification of spin-split electron bands quickly followed their discovery \cite{fedchenkoObservationTimereversalSymmetry2024, krempaskyNature2024, OsumiPRB2024, reimers2024direct, zengObservationSpinSplitting2024, dingLargeBandsplittingWave2024}.

In contrast to collinear magnets such as altermagnets, noncollinear and noncoplanar magnets can host itinerant spin-split electron bands with an unconventional odd parity, e.g., $p$-wave, symmetry induced by an isotropic sd-spin interaction \cite{martin_prb_12, Hayami-Kusunose-SpontaneousAntisymmetricSpin-2020, Hayami-Kusunose-BottomupDesignSpinsplit-2020, Hayami-Hayami-MechanismAntisymmetricSpin-2022, KudasovPRB2024, hellenesUnconventionalPwaveMagnets2024}. The presence of a composite $\mathcal{T}\bm{\tau}$ symmetry protects the $p$-wave spin polarization in noncentrosymmetric and noncollinear magnets~\cite{KudasovPRB2024, hellenesUnconventionalPwaveMagnets2024}; here, $\mathcal{T}$ is the time-reversal operator and $\bm{\tau}$ denotes a translation, typically by half a unit cell. This symmetry makes $p$-wave magnets distinct from general helimagnets, where $p$-wave spin polarization can also be observed in specific cases~\cite{Chen-Ramires:2022, Hayami-Hayami-MechanismAntisymmetricSpin-2022, KudasovPRB2024, Mayo-Onose:2024}. While $p$-wave magnets were classified and material candidates, such as Mn$_3$GaN and CeNiAsO, were proposed in Ref.~\cite{hellenesUnconventionalPwaveMagnets2024}, the potential spintronics applications and even effective models of $p$-wave magnets remain unexplored.

In this Letter, we fill this knowledge gap and formulate an effective low-energy, analytical Hamiltonian for unconventional $p$-wave magnets. It is based on a minimal magnetic lattice featuring $p$-wave spin-polarized bands protected by $\mathcal{T}\bm{\tau}$ symmetry. As an application of the model and a manifestation of the $p$-wave nature of unconventional magnets, we predict a large tunneling magnetoresistance (TMR) in junctions of $p$-wave magnets with the qualitatively different dependence on the relative orientation of the spin-spin Fermi surfaces: unlike $d$-wave altermagnets, the antiparallel configuration of spins needed for TMR is achieved by rotating the $p$-wave magnet by $\pi$ rather than $\pi/2$. The $p$-wave magnet also filters injected spin currents from a different material. Finally, we show that while $p$-wave magnets feature an equilibrium spin current, applying an electric field also produces a nonequilibrium transport spin current to the second order in the applied field. The TMR and spin currents provide effective means to detect and utilize $p$-wave magnets in spintronics.

\textit{Symmetry groups and minimal models for $p$-wave antiferromagnets.---}
Spin space groups classify collinear and noncollinear magnets with a negligible coupling between the crystal lattice and spin spaces \cite{brinkman1966theory, corticelli2022spin, xiao2023spin, chen2023enumeration, Cheong-Huang:2024, Singh-Guo:2024}. Electron spin splitting can still arise in the absence of broken $\mathcal{P}\mathcal{T}$ symmetry, where $\mathcal{P}$ is the inversion operator. Altermagnets have even parity spin-split electron bands consistent with rotational symmetry. This rotational symmetry prohibits net magnetization and ensures the altermagnet is fully compensated; it is a defining property of altermagnets. In contrast, the absence of magnetization in unconventional $p$-wave magnets is enforced by a composite $\mathcal{T}\bm{\tau}$ symmetry. Here, we consider spin-polarized electron bands with a $p$-wave character mediated by spin-spin interactions, protected by this $\mathcal{T}\bm{\tau}$ symmetry \cite{KudasovPRB2024}. We take this as the definition of unconventional $p$-wave magnets \cite{hellenesUnconventionalPwaveMagnets2024}.

\begin{figure}[ht!]
\centering
\begin{subfloat}[\label{subfig:pWaveLattice}]{\includegraphics[width=0.99\columnwidth]{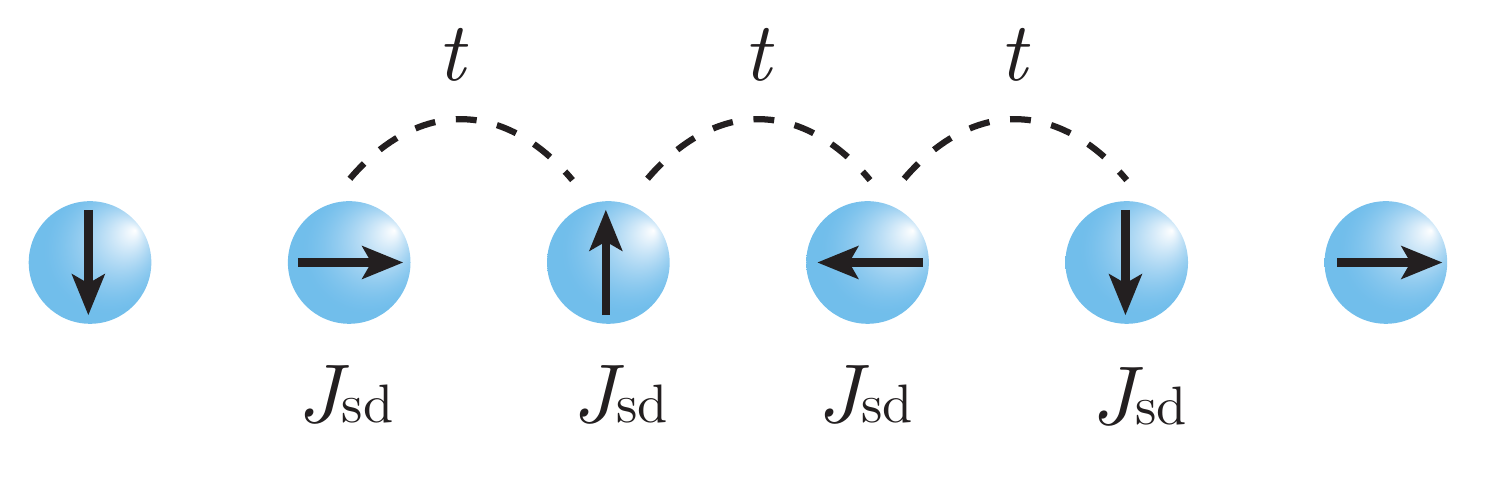}}
\end{subfloat}
\begin{subfloat}[\label{subfig:electronBands}]{\includegraphics[width=0.99\columnwidth]{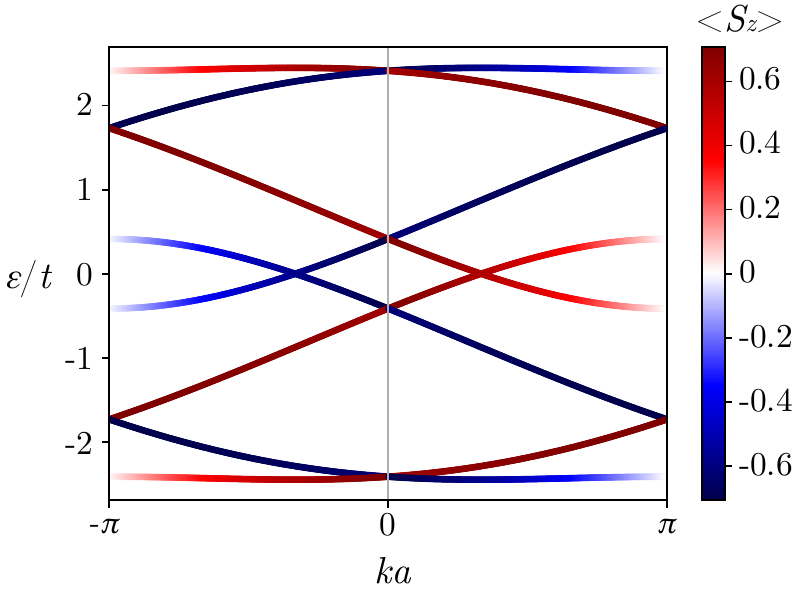}}
\end{subfloat}
\caption{\label{fig:minimalModel} (a) A minimal magnetic lattice for $p$-wave magnetism. The nearest neighbor sites are coupled by hopping $t$, and the itinerant electrons interact with the localized spins by the sd interaction $J_{\mathrm{sd}}$. (b) The electron band structure with spin polarization $\left<S_z\right>$ of the bands in units of $\hbar/2$ shown in red and blue. The bands are plotted for parameters $J/t=1$ and $\mu/t=0$ in Eq.~(\ref{latticeModelHamiltonian}).
}
\end{figure}

As an example of a coplanar $p$-wave magnet, we propose the minimal lattice model shown in Fig. \ref{subfig:pWaveLattice}. The lattice is centrosymmetric. However, the magnetic ordering breaks both $\mathcal{P}$ and $\mathcal{P}\mathcal{T}$ symmetries. Since the spin rotates by $\pi/2$ between each lattice site, the proposed model belongs to the class of helimagnets (cycloidal magnets). However, the $\mathcal{T}\bm{\tau}$ symmetry, crucial for unconventional $p$-wave magnets, is not a general property of helimagnets. For example, helimagnets, where the spin rotates by $2\pi/(2n+1)$ with $n\in \mathds{Z}$, have no $\mathcal{T}\bm{\tau}$ symmetry and are not guaranteed to have a $p$-wave spin polarization for the itinerant electrons \cite{KudasovPRB2024}. While a $p$-wave polarization can be protected by other accidental symmetries, these symmetries are, in contrast to the $\mathcal{T}\bm{\tau}$ symmetry in unconventional $p$-wave magnets, not robust in the presence of spin-orbit coupling.

The minimal $p$-wave magnetic lattice in Fig.~\ref{subfig:pWaveLattice} is described by the following tight-binding model:
\begin{align}\nonumber
    H_e =& -t\sum_{\left<i,j\right>,\sigma}  c_{i,\sigma}^{\dagger} c_{j,\sigma} -\mu \sum_{i, \sigma}  c_{i,\sigma}^{\dagger} c_{i,\sigma}
  \\
  &- J_{\mathrm{sd}}\sum_{i, \sigma, \sigma'} \bm{S}_i\cdot c_{i,\sigma}^{\dagger}\bm{\sigma}_{\sigma \sigma'} c_{i,\sigma'},
    \label{latticeModelHamiltonian}
\end{align}
where $c_{i,\sigma}^{(\dagger)}$ is the electron (creation) annihilation operator for an electron with spin $\sigma$ at site $i$. The parameter $t$ corresponds to hopping between nearest neighbors $\left<i,j\right>$, $\mu$ is the chemical potential, and $J_{\mathrm{sd}}$ is the isotropic sd coupling between a localized spin and itinerant electrons. The tight-binding Hamiltonian features $p$-wave spin-polarized bands, as shown in Fig. \ref{subfig:electronBands}. The spin polarization of the electron bands is collinear and perpendicular to all localized spins, such that $\left<S_x\right>=\left<S_y\right>=0$.

\begin{figure}[ht!]
\centering
\begin{subfloat}[\label{subfig:effectiveModelLattice}]{\includegraphics[width=0.99\columnwidth]{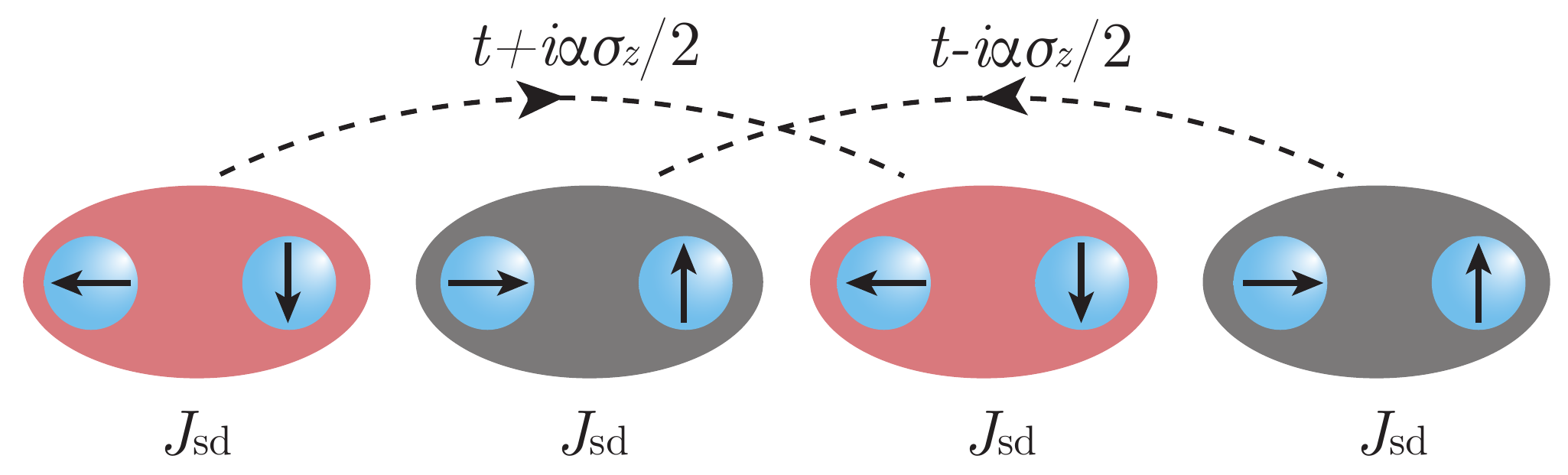}}
\end{subfloat}
\begin{subfloat}[\label{subfig:polarizedBandsEffective}]{\includegraphics[width=0.99\columnwidth]{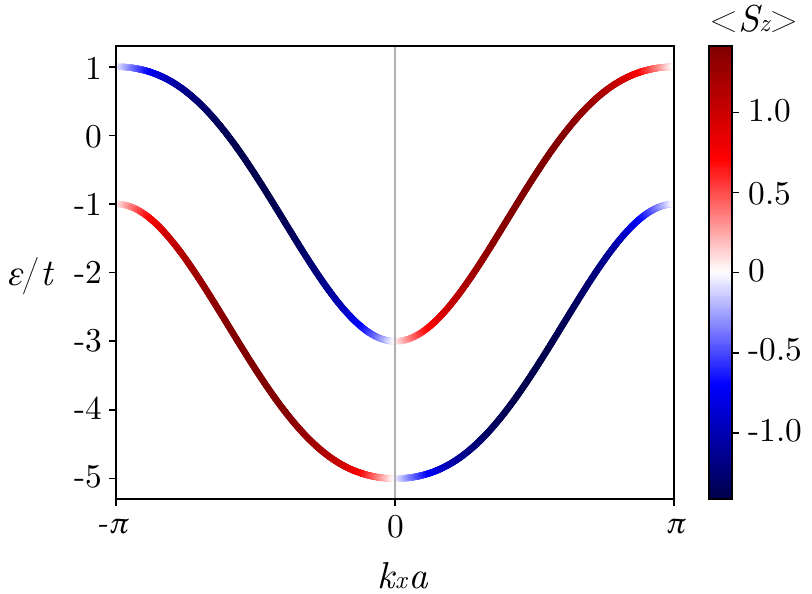}}
\end{subfloat}
\caption{\label{fig:effectiveModel} (a) A lattice model consistent with a 1D version of the effective model in Eq. \eqref{effectiveModel}. The model has two sectoral degrees of freedom, as shown in red and gray. (b) The resulting electron bands with spin polarization $\left<S_z\right>$ in units of $\hbar/2$. The bands are calculated via Eq.~\eqref{spectrum-effective} with parameters $J_{\mathrm{sd}}/t = \alpha/t = 1$ and $\mu/t=0$.
}
\end{figure}

\textit{Phenomenological model for itinerant electrons in $p$-wave magnets.---}
Motivated by the minimal lattice model in Fig. \ref{subfig:pWaveLattice}, we propose a 2D effective model for itinerant electrons in a $p$-wave magnet
\begin{align}
    \nonumber
    H_{\mathrm{eff}} =& -\left\{2t \left[\cos{(k_x a)} + \cos{(k_y a)}\right] +\mu \right\}\sigma_0 \otimes \tau_0 \\ \nonumber
    &+ \left[\alpha_x \sin{(k_x a)} + \alpha_y \sin{(k_y a)}\right] \sigma_{z'} \otimes \tau_0 \\ &+ J_{\mathrm{sd}} \sigma_{x'} \otimes \tau_z.
    \label{effectiveModel}
\end{align}
Here, $\sigma_i$ and $\tau_i$ are Pauli matrices acting on the spin space and the degree of freedom spanned by the two sectors depicted as red and gray ovals in Fig. \ref{subfig:effectiveModelLattice}. The sectors host localized spins that are related by the mentioned $\mathcal{T}\bm{\tau}$ symmetry. In particular, $\mathcal{T}$ reverses the spin and $\bm{\tau}$ exchanges the sectors $\bm{\tau}\tau_z \rightarrow -\tau_z$. This is a general feature of unconventional $p$-wave magnets, not only the minimal lattice in Fig. \ref{subfig:pWaveLattice}.
The spin coordinates are primed to emphasize that they are decoupled from the crystal coordinates. While we set them equal for simplicity, different magnetic textures in 2D can be modeled by rotating these spin coordinates. For example, for a magnetic texture that avoids induced in-plane electric polarization~\cite{Mostovoy-FerroelectricitySpiralMagnets-2006}, one should replace $z'\to x$ and $x'\to y$; this does not affect the p-wave symmetry and transport properties presented below.
Here, $t$ denotes the intrasectoral hopping shown in Fig.~\ref{fig:effectiveModel}, the net sd coupling within a sector is given by $J_{\mathrm{sd}}$, and $a$ is the length of the primitive lattice vectors. Within the effective model, each site hosts two localized spins. Consequently, the model becomes effectively collinear by a unitary spin-space rotation. For this reason, we add spin-dependent hopping with strengths $\alpha_x$ and $\alpha_y$ to recover the noncollinear magnetic nature of the model. The subscripts $x$ and $y$ refer to the crystal directions and follow from a natural two-dimensional generalization of the model. We neglect intersectoral hopping for convenience since it does not affect the defining properties of the $p$-wave magnet; see the Supplemental Material~\cite{SM} for the discussion. Consequently, the effective model is a $4\times 4$ Hamiltonian block-diagonal in the sectoral degree of freedom.

The four electron bands are two-fold degenerate with energies given by
\begin{align}
\label{spectrum-effective}
    \nonumber
    \varepsilon_{\pm} =& -2t \left[\cos{(k_x a)} + \cos{(k_y a)}\right] - \mu \\ &\pm \sqrt{J_{\mathrm{sd}}^2 + \big[\alpha_x\sin{(k_x a)} + \alpha_y\sin{(k_y a)}\big]^2},
\end{align}
where the band splitting is determined by $J_{\mathrm{sd}}$. Each of the degenerate bands is spin polarized, as exemplified in Fig. \ref{subfig:polarizedBandsEffective}. The combined spin polarization of these two doubly degenerate bands along the crystal $k_x$ axis is given by
\begin{align}
    \left<S_z\right> = \pm \frac{2\alpha_x \sin{(k_x a)}}{\sqrt{J_{\mathrm{sd}}^2 + \left[\alpha_x\sin{(k_x a)}\right]^2}},
\end{align}
in units of $\hbar/2$ whereas the spin polarizations $\left<S_x\right>$ and $\left<S_y\right>$ cancel between the degenerate copies. The inclusion of the intersectoral hopping lifts the twofold degeneracy leaving only nonvanishing $\left<S_z\right>$~\cite{SM}.

While the model (\ref{spectrum-effective}) is topologically trivial, it can be extended to include nontrivial topology. For example, the inclusion of the spin-dependent hopping modulates $J_{\rm sd} \to J_{\rm sd} +J_{1} \cos{(k_xa)}$ in Eq.~(\ref{effectiveModel}), which allows for nonzero winding numbers in each of the sectors at, e.g., $J_{1} \cos{(k_x a)} \gg J_{\mathrm{sd}}$~\cite{SM}.

In contrast to altermagnets without spin-orbit coupling, unconventional $p$-wave magnets feature only partial spin polarization. This is an integral property of noncollinear magnets. Notably, a strong sd coupling $J_{\mathrm{sd}}$ induces a large spin splitting but reduces the spin-polarization of the bands, making $J_{\mathrm{sd}}$ a trade-off quantity for the potential spintronics properties of unconventional $p$-wave magnets.

\begin{figure}[ht!]
\centering
\begin{subfloat}[\label{subfig:TMR-model}]{\includegraphics[width=0.95\columnwidth]{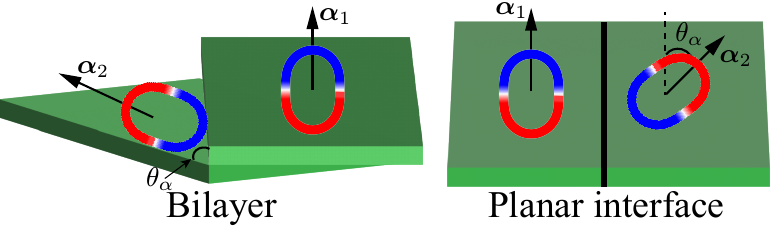}}\end{subfloat}
\begin{subfloat}[\label{subfig:TMR-bilayer}]{\includegraphics[width=0.9\columnwidth]{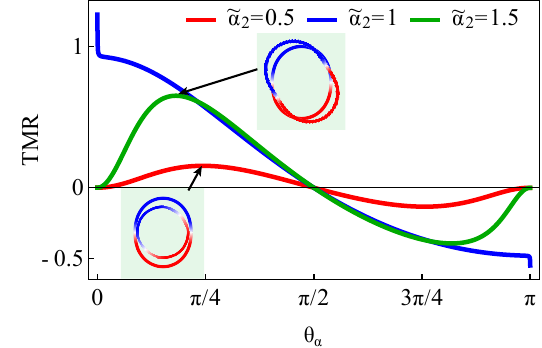}}
\end{subfloat}
\begin{subfloat}[\label{subfig:TMR-planar}]{\includegraphics[width=0.9\columnwidth]{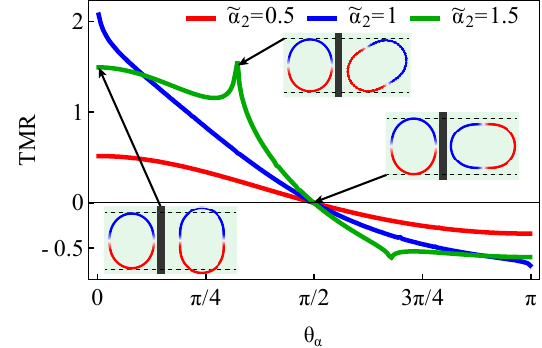}}
\end{subfloat}
\label{fig:TMR}
\caption{
(a) Model setups of the bilayer and the planar junction of $p$-wave magnets.
(b) TMR as a function of the relative angle $\theta_{\alpha}$ between the spin-splitting vectors $\bm{\alpha}_1$ and $\bm{\alpha}_2$ in two layers of the bilayer.
(c) The same as (b) in a planar junction, where we fix $\bm{\alpha}_1$ to be parallel to the interface. The insets schematically depict the Fermi surfaces at $\tilde{\alpha}_2=0.5$ and $\tilde{\alpha}_2=1.5$ (b) as well as $\tilde{\alpha}_2=1.5$ (c). The conductance is calculated in the linear response regime. We use $\tilde{J} = J_{\rm sd}/\mu$ and $\tilde{\bm{\alpha}} = 2m \bm{\alpha}'/k_F$ with $k_F=\sqrt{2m\mu}$ being the Fermi wave vector and set $\tilde{J}_1=\tilde{J}_2=\tilde{\alpha}_1=1$.
}
\end{figure}

\textit{Tunneling magnetoresistance.---}
As a potential experimental probe of $p$-wave magnets, we suggest the TMR. We calculate the TMR in two model setups involving 2D $p$-wave magnets: (i) a bilayer and (ii) a planar interface, see Fig.~\ref{subfig:TMR-model}. To make an analytical advance, we consider a low-energy expansion of the Hamiltonian (\ref{effectiveModel}):
\begin{equation}
\label{7-tmr-H-i-lin}
H_{\eta}(\mathbf{k}) = \left(\frac{k^2}{2m} -\mu\right) \sigma_0 + \left(\bm{\alpha}'\cdot\mathbf{k}\right)\sigma_{z} + \eta J_{\rm sd} \sigma_x,
\end{equation}
where $k=\sqrt{k_x^2+k_y^2}$, $m=1/(2ta^2)$ is the effective mass, $\bm{\alpha}' = a\bm{\alpha}$, and $\eta=\pm$ corresponds to the sectoral degree of freedom. Whereas Eq. (\ref{7-tmr-H-i-lin}) describes a $p$-wave spin polarization even with $J_\text{sd}=0$, it is essential to note that such a model would not account for spin-flip processes which inevitably are present in actual inhomogeneous and noncollinear magnets that were numerically predicted to host a $p$-wave spin polarization~\cite{hellenesUnconventionalPwaveMagnets2024}.
Importantly, such processes are included in our model via the $J_\text{sd}$ term.

The TMR is defined as~\cite{Tsymbal-LeClair:2003}
\begin{equation}
\label{TMR-def}
\mathrm{TMR}= \frac{G(\theta_{\alpha})-G(\theta_{\alpha}+\pi)}{G(\theta_{\alpha}+\pi)},
\end{equation}
where $G(\theta_{\alpha})$ is the differential conductance between the $p$-wave magnets calculated at vanishing bias as a function of the relative angle $\theta_{\alpha}$ between the splitting vectors $\bm{\alpha}_{1,2}$ in the magnets~\footnote{Similar to the helical orientation in helimagnets~\cite{Masell-Nagaosa:2020}, the direction of the spin-splitting vector $\bm{\alpha}$ may be controllable via current pulses.}. The $p$-wave character of the magnets is evident from the periodicity of the TMR. Unlike $d$-wave magnets (altermagnets) studied in Refs.~\cite{SmejkalPRX22Hellenes,shao2021spin}, the spin flip necessary for the TMR in the present $p$-wave case is achieved when the magnets are rotated by $\pi$ rather than $\pi/2$ and the TMR vanishes when the relative angle between the magnets is $\pi/2$ rather than $\pi/4$. The angular dependence of the TMR provides a direct way to identify the $p$-wave spin splitting of the magnet.

To calculate the differential conductance that enters the TMR~(\ref{TMR-def}), we use the standard tunneling Hamiltonian approach; see the End Matter and the Supplemental Material~\cite{SM}. The TMR in the bilayer and planar junctions are shown in Figs.~\ref{subfig:TMR-bilayer} and \ref{subfig:TMR-planar}, respectively. Both junctions allow for a sizable TMR that originates from the mismatch of the spin polarizations in the magnets. The TMR has a larger magnitude and is less sensitive to the mismatch of the Fermi surfaces in the planar junction, cf. Figs.~\ref{subfig:TMR-bilayer} and \ref{subfig:TMR-planar}. Furthermore, it shows additional features originating from the interplay of the spin polarization and the overlap of the Fermi surfaces in different magnets. For instance, the sharp maximum at $\theta_{\alpha}=0$ for equivalent magnets in the bilayer system, see the blue line in Fig.~\ref{subfig:TMR-bilayer}, originates from the complete overlap of the Fermi surfaces that is drastically reduced by any misalignment, e.g., at $\theta_{\alpha}>0$; the width of the maximum depends on the broadening. In the planar junction, a peak occurs when the Fermi surfaces on both sides of the junction match, see the green line in Fig.~\ref{subfig:TMR-planar}. The nonmonotonic dependence of the TMR is related to the shape of the Fermi surfaces and can also occur for equivalent magnets in the presence of, e.g., intersectoral hopping.

Combining a $p$-wave magnet with a regular metal allows for spin filtering through different conductances for spin-up and spin-down electrons. The filtering is observed for the orientation of the interface that does not respect the combined translation and time-reversal symmetry of the $p$-wave magnet, i.e., $\bm{\alpha}$ should have a component perpendicular to the interface. It is also worth noting that, in the same setup, antisymmetric magnets with two fully spin-polarized Fermi surfaces, reminiscent of a 1D spin-orbit coupled wire, do not allow for spin filtering~\cite{SM} allowing one to distinguish unconventional $p$-wave magnets in spin-transport measurements.

\textit{Spin conductivity.---}
As their altermagnetic cousins, $p$-wave magnets offer a viable route to generate spin currents. For inhomogeneous magnetic textures such as the present system, a spin current exists in equilibrium without any applied bias \cite{katsura_prl_05}. This can be intuitively understood based on the Fermi surface of $p$-wave magnets where carriers with opposite momentum have opposite spin, thus creating a net spin flow along the spin-splitting vector $\bm{\alpha}$. Such an exchange spin current should be distinguished from a transport (nonequilibrium) spin current that can be induced by an external bias \cite{Rashba2003PRB}. By using the semiclassical approach with the collision integral in the relaxation-time approximation, we calculate the first- and second-order response to an electric field in $p$-wave magnets; see the Appendix and the Supplemental Material~\cite{SM} for details. Since $p$-wave magnets enjoy the $\mathcal{T}\bm{\tau}$ symmetry, only even-order (odd-order) in the electric field spin (electric) current response coefficients are nonzero. For simplicity, we consider the limit of small Fermi energy $|\mu|<|J_{\rm sd}|$ where only the lowest-energy bands, see Fig.~\ref{subfig:polarizedBandsEffective}, contribute to electric and spin transport; this allows us to neglect inter-band effects.

In the linear-response regime, we find an anisotropic electric conductivity tensor with a vanishing spin conductivity. The anisotropy of the former is the manifestation of the shape of the Fermi surface: electrons moving along $\bm{\alpha}$ have lower velocity.

The electric-field-induced spin current is obtained in the second-order response, $J_i^{(\sigma)} = \chi_{ijl}^{(\sigma)} E_jE_l$; the second-order electric current response vanishes. The symmetry of the Fermi surface in the direction perpendicular to $\bm{\alpha}$ permits only $\chi_{\parallel \parallel \parallel}^{(\sigma)}$, $\chi_{\parallel \perp \perp}^{(\sigma)}$, and $\chi_{\perp  \perp \parallel}^{(\sigma)}=\chi_{\perp \parallel \perp}^{(\sigma)}$, where $\parallel$ ($\perp$) denotes the component parallel (perpendicular) to $\bm{\alpha}$.

\begin{figure}[ht!]
\centering
{\includegraphics[width=0.9\columnwidth]{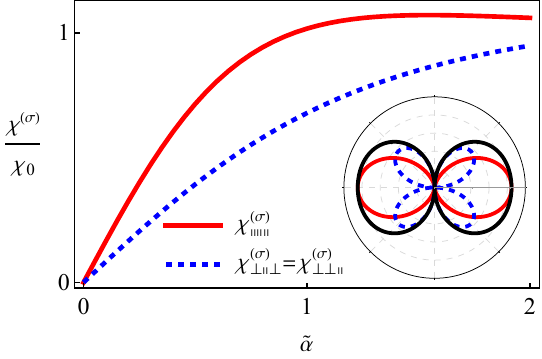}}
\caption{The nontrivial components of the second-order spin response tensor $\chi_{ijl}^{(\sigma)}$ as a function of the spin-splitting parameter $\tilde{\alpha}$. The inset shows the dependence of the spin current components $j_{\sigma,\parallel}$ (solid red line), $j_{\sigma,\perp}$ (dashed blue line), and the magnitude of the current (solid black line) on the angle between $\mathbf{E}$ and $\bm{\alpha}$ at $\tilde{\alpha} =1$. We use $\chi_0 = e^3 \tau^2 k_F/m$ with $\tau$ being relaxation time and set $\tilde{J}=1.5$.
}
\label{fig:spin-current}
\end{figure}

We show the corresponding response coefficients in Fig.~\ref{fig:spin-current} as functions of the spin-splitting parameter $\tilde{\alpha}$. Only $\chi_{\parallel \parallel \parallel}^{(\sigma)}$ and $\chi_{\perp  \perp \parallel}^{(\sigma)}=\chi_{\perp \parallel \perp}^{(\sigma)}$ are finite. This can be understood from the following qualitative picture: an electric field along $\bm{\alpha}$ leads to an imbalance between the populations at $\pm\mathbf{k}\parallel \mathbf{E}$. Because of the spin polarization of the bands, see Fig.~\ref{subfig:polarizedBandsEffective}, this imbalance directly translates into the spin polarization and, as a result, into the spin current. In this picture, no spin imbalance should occur at $\mathbf{E}\perp\bm{\alpha}$, hence $\chi_{\parallel \perp \perp}^{(\sigma)}=0$. We expect that the $\mathbf{E}$-induced spin current gives rise to a spin accumulation.

\textit{Concluding remarks.---}
Our effective 2D low-energy model for unconventional $p$-wave magnets is given in Eq.~(\ref{effectiveModel}). It is motivated by the presence of a $\mathcal{T}\bm{\tau}$ symmetry exemplified by a four-site lattice model, see Fig.~\ref{subfig:pWaveLattice}, and features two doubly degenerate bands with partial $p$-wave spin polarization shown in Fig.~\ref{fig:effectiveModel}. This symmetry distinguishes the unconventional $p$-wave magnets from other systems with antisymmetric spin splitting. Inclusion of intersectoral hopping lifts the degeneracy but preserves the $p$-wave symmetry for the $\langle S_z\rangle$ polarization.

We used the effective model to describe the spin transport properties in heterostructures exemplified by the TMR, spin filtering, and spin current responses. The characteristic dependence of the TMR on the relative angle between the spin-splitting vectors $\bm{\alpha}$ in the magnets composing the heterostructures provides an experimentally viable way to identify $p$-wave magnets: the spin-flip necessary for the TMR is achieved by rotating one of the magnets by $\pi$ rather than $\pi/2$ as in $d$-wave altermagnets. The relatively large magnitude of the TMR $\sim 100-200~\%$ makes $p$-wave magnets promising for spintronic applications. Spin filtering is achieved in a junction of a $p$-wave magnet and a metal. This effect is not guaranteed for other systems with antisymmetric spin splitting and, hence, may serve as an indicator of unconventional $p$-wave magnets. Similar to $d$-wave magnets~\cite{GonzalezPRL2021Splitter}, $p$-wave magnets allow for a spin current accompanied by an electric current. However, in the model at hand, there is no spin-splitter effect~\cite{GonzalezPRL2021Splitter}, where the spin current is perpendicular to the applied
electric field.

The proposed model can readily address the transport properties of superconducting junctions and interaction effects in $p$-wave magnets. Indeed, as in helical and cycloidal magnets~\cite{Rex-Mirlin:2020}, the nontrivial spin texture of our model realizes synthetic spin-sectoral coupling, which could allow for Majorana bound states. Therefore, our model provides a playground to study the interplay of unconventional magnetism, superconductivity, and, after a model extension, topology~\footnote{The topology of electron band structures in helimagnets was recently studied in Ref.~\cite{KudasovPRB2024}.}.

\begin{acknowledgments}
{\it Acknowledgments.} We thank Jeroen Danon for useful comments.
The Research Council of Norway (RCN) supported this work through its Centres of Excellence funding scheme, project number 262633, "QuSpin", and RCN project number 323766.
\end{acknowledgments}

%


\pagebreak
\begin{widetext}
\begin{center}
\textbf{End Matter}
\end{center}
\end{widetext}
\setcounter{equation}{0}
\setcounter{figure}{0}
\setcounter{table}{0}
\makeatletter
\renewcommand{\theequation}{A\arabic{equation}}
\renewcommand{\thefigure}{A\arabic{figure}}

\textit{Appendix A: TMR.---}
The TMR measures how the differential conductance $G=dI/dV$ between two magnets changes when the relative magnetization direction of the magnets flips from parallel to antiparallel, and is defined in Eq.~(6) in the main text. The differential conductance reads~\cite{Levitov-Shytov:book, Mahan:book-2013}
\begin{equation}
\label{2-tmr-G}
G = 4e\pi^3 \sum_{\mathbf{k},\mathbf{p}} |T_{\mathbf{p},\mathbf{k}}|^2\,  \mbox{Tr}{\left\{\mbox{Im}{\left[G_{1}(0;\mathbf{p})\right]} \mbox{Im}{\left[G_{2}(0;\mathbf{k})\right]} \right\}}.
\end{equation}
We assumed the zero-temperature limit and considered the limit of small voltage biases between the magnets $|eV|\ll \mu$ with $\mu$ being the Fermi energy. Furthermore, $G_{i=1,2}(\omega;\mathbf{k})$ is the retarded Green's function of the $i$th magnet, and $T_{\mathbf{p},\mathbf{k}}$ is the tunneling coefficient. The trace runs over spin and sectoral degrees of freedom.

We consider two setups: (i) a bilayer with $|T_{\mathbf{p},\mathbf{k}}|^2 = T_0^2 \delta_{\mathbf{p},\mathbf{k}}$ which preserves in-plane momenta and (ii) a planar junction with $|T_{\mathbf{p},\mathbf{k}}|^2 = T_0^2 \delta_{p_{\parallel},k_{\parallel}}/L$ which preserves only the momentum along the interface with the size $L$.

The retarded Green's function used in Eq.~(\ref{2-tmr-G}) is $G_{i}(\omega;\mathbf{k}) = \left[\omega +i0^{+} - H_i(\mathbf{k})\right]^{-1}$. In the absence of the intersectoral hopping, the Hamiltonian $H_i(\mathbf{k})$ becomes block diagonal in the sectoral space. In calculating the imaginary parts of the Green's functions in the conductance (\ref{2-tmr-G}), we use the Sokhotski–Plemelj theorem $\IM{1/(x+i0^{+})} = -\pi \df{x}$. Then, the summation (integration) over $k$ and, in the case of a planar junction, $p_{\parallel}$ can be carried out analytically. The remaining integrals are calculated numerically where we treat the $\delta$ functions as Lorentzians with the half-width $\Gamma$. In our numerical calculations, $\Gamma/\mu=10^{-3}$. For the planar interface, we fix the spin-splitting vector $\bm{\alpha}$ along the interface in one of the magnets and only vary the direction of the spin-splitting vector in the other magnet.

\textit{Appendix B: Spin filtering.---}
The spin filtering effect is considered for a planar interface between a $p$-wave magnet and a regular metal. The corresponding Hamiltonian is
\begin{widetext}
\begin{equation}
\label{8-i-H-lin}
H_{\eta}(x,k_y) =\left(-\frac{\nabla^2_{x}}{2m} +\frac{k_{y}^2}{2m} -\mu\right) \sigma_0 + \left[-\frac{i}{2}\left\{\alpha_x(x),\nabla_x\right\} +\alpha_y(x) k_{y}\right]\sigma_{z} + \eta J_{\rm sd}(x) \sigma_x +U(x)\sigma_0,
\end{equation}
\end{widetext}
where $\eta=\pm$ denotes the sectoral degree of freedom, $\bm{\alpha}(x) = \bm{\alpha} \Theta{(x)}$, and $J_{\rm sd}(x) = J_{\rm sd} \Theta{(x)}$ with $\Theta{(x)}$ being the unit-step function; the barrier at the interface is modeled as $U(x)=U\delta(x)$~\cite{BTK:1982}. For definiteness, we fixed the $x$ and $y$ axes to be perpendicular and parallel to the interface, respectively.

The differential conductance per spin projection $s$ in the metallic part of the junction is determined by
\begin{equation}
\label{8-i-G-e-0}
G_{i;s}(V) = -e^2 L^3 \sum_{\eta=\pm} \int \frac{dk_y}{(2\pi)^2} \int d\varepsilon \frac{\partial k_x}{\partial \varepsilon} j_{i;\eta,s} f'(\varepsilon -eV),
\end{equation}
where $i$ corresponds to the electric ($i=el$) and spin ($i=\sigma$) conductances, $k_x = \sqrt{2m\varepsilon -k_y^2}$ follows from the dispersion relation of the metal [see Eq.~(\ref{8-i-H-lin})], and prime denotes the derivative of the distribution function $f$ with respect to the argument. The current used in Eq.~(\ref{8-i-G-e-0}) is defined as
\begin{equation}
\label{8-i-je-def}
j_{i;\eta,s} = \frac{1}{2L^2}\RE{\psi_{\eta,s}^{\dag} \hat{j}_{i;\eta,s} \psi_{\eta,s}}.
\end{equation}
Here, the electric and spin current operators in the metal are $\hat{j}_{i=el;\eta,s} =\overleftrightarrow{\hat{v}}_x$ and $\hat{j}_{i=\sigma;\eta,s} = \sigma_{z} \overleftrightarrow{\hat{v}}_x$, respectively, and arrows indicate that the velocity operator $\hat{v}_x = -i\nabla_{x}/m$ acts on both $\psi_{\eta,s}^{\dag}$ and $\psi_{\eta,s}$. The latter are found by matching the incident, reflected, and transmitted states and their derivatives at the interface.

\textit{Appendix C: Spin conductivity.---}
In the analysis of electric and spin transport properties of bulk $p$-wave magnets, we focus on the low-energy model and apply the semiclassical kinetic approach. The distribution function $f$ satisfies the following kinetic equation:
\begin{equation}
\label{8-kin-eq}
-e\mathbf{E}\cdot\partial_{\mathbf{k}} f = -\frac{f-f_{eq}}{\tau}.
\end{equation}
Here, $\mathbf{E}$ is the electric field, $f_{eq}=1/\left[e^{(\varepsilon_{\mathbf{k}}-\mu)/T}+1\right]$ is the equilibrium distribution function, $\varepsilon_{\mathbf{k}}$ is the dispersion relation, $T$ is temperature, and $\tau$ is the relaxation time. In writing Eq.~(\ref{8-kin-eq}), we assumed a uniform and time-independent electric field.

For $eE \tau v_F/\mu\ll1$ with $v_F$ being the Fermi velocity, we solve Eq.~(\ref{8-kin-eq}) perturbatively, $f = f^{(0)} +f^{(1)} +f^{(2)} +\ldots$. Here, $f^{(0)} =f_{eq}$ and $n>0$ terms are $f^{(n)} = e \tau \left(\mathbf{E}\cdot\partial_{\mathbf{k}}\right) f^{(n-1)}$.

The electric current density $\mathbf{j}_{el}$ in each order of the perturbation theory is defined as
\begin{equation}
\label{8-kin-j-n}
\mathbf{j}_{el}^{(n)} = -e \int \frac{d\mathbf{k}}{(2\pi)^2} \mathbf{v} f^{(n)},
\end{equation}
where $\mathbf{v} = \partial_{\mathbf{k}} \varepsilon_{\mathbf{k}}$ is the group velocity. In the spin current density $\mathbf{j}_{\sigma}$, one should replace $\mathbf{v} \to \left\langle \hat{\mathbf{j}}_{\sigma} \right\rangle$ with $\left\langle \hat{\mathbf{j}}_{\sigma} \right\rangle$ being the mean value of the spin current operator. By using the continuum model and considering only the lowest energy band, $\varepsilon_{\mathbf{k}} = k^{2}/(2m) -\sqrt{J^2 +\left(\bm{\alpha}\cdot\bm{k}\right)^2}$, we obtain
\begin{equation}
\label{8-kin-vz-Sz-continuity}
\mathbf{v} = \frac{\mathbf{k}}{m} +\bm{\alpha} S_z \quad \quad  \hat{\mathbf{j}}_{\sigma} = \frac{\mathbf{k}}{m} \sigma_z +\bm{\alpha}.
\end{equation}
The mean value $\left\langle \hat{\mathbf{j}}_{\sigma} \right\rangle$ is given by Eq.~(\ref{8-kin-vz-Sz-continuity}) with $\sigma_z\to \left\langle S_z\right\rangle = - (\bm{\alpha}\cdot \mathbf{k})/\sqrt{J^2 + (\bm{\alpha}\cdot\mathbf{k})^2}$.

The group velocity $\mathbf{v}$ and the mean value of the spin current operator $\left\langle \hat{\mathbf{j}}_{\sigma} \right\rangle$ are odd and even in momentum, respectively.

Assuming the zero-temperature limit, the first-order in electric field contribution to the electric current is
\begin{eqnarray}
\label{8-kin-j-1-el}
\mathbf{j}_{el}^{(1)} = e^2 \tau \!\int \frac{d\mathbf{k}}{(2\pi)^2} \mathbf{v} \left(\mathbf{E}\cdot\mathbf{v}\right) \df{\varepsilon_{\mathbf{k}} -\mu}
\end{eqnarray}
with $\mathbf{v} \left(\mathbf{E}\cdot\mathbf{v}\right) \to \left\langle \hat{\mathbf{j}}_{\sigma} \right\rangle \left(\mathbf{E}\cdot\mathbf{v}\right)$ in the spin current. Since $\varepsilon_{-\mathbf{k}}=\varepsilon_{\mathbf{k}}$ due to the $\mathcal{T}\bm{\tau}$ symmetry of $p$-wave magnets, the integrands in the first-order electric and spin currents are even and odd, respectively, functions of momenta. Therefore, the first-order spin current vanishes.

A nonzero electric current response is quantified by anisotropic longitudinal and vanishing transverse electric conductivities, which, for small $\tilde{\alpha} =2m \alpha/k_F$, read
\begin{eqnarray}
\label{8-kin-i-2-sigma-xx-el-app}
\sigma_{\parallel}^{(el)} &\approx& \sigma_0  \left(1+|\tilde{J}| \right) \left(1 - \frac{7 \tilde{\alpha}^2}{16|\tilde{J}|}\right),\\
\label{8-kin-i-2-sigma-yy-el-app}
\sigma_{\perp}^{(el)} &\approx& \sigma_0  \left(1+|\tilde{J}| \right) \left(1 +\frac{3\tilde{\alpha}^2}{16|\tilde{J}|}\right).
\end{eqnarray}
Here, $\tilde{J} =J/\mu$ and $\sigma_0 = e^2\tau \mu/\pi$.

The decrease of the conductivity along the direction of the spin splitting $\tilde{\bm{\alpha}}$, i.e., $\sigma_{\parallel}^{(el)}$, can be understood from the band structure: with the rise of $\tilde{\alpha}$, the energy dispersion becomes more shallow along $\tilde{\bm{\alpha}}$ leading to the decrease of the group velocity; the trend is opposite in the direction perpendicular to $\tilde{\bm{\alpha}}$, i.e., for $\sigma_{\perp}^{(el)}$. Note that the anisotropy of the electric conductivity is the manifestation of the Fermi surface anisotropy and does not reflect the spin polarization of $p$-wave magnets.

The second-order electric current is
\begin{equation}
\label{8-kin-j-2-el}
\mathbf{j}_{el}^{(2)} = e^3 \tau^2 \int \frac{d\mathbf{k}}{(2\pi)^2} \mathbf{v} \left(\mathbf{E}\cdot\partial_{\mathbf{k}}\right) \left(\mathbf{E}\cdot\mathbf{v}\right) \df{\varepsilon_{\mathbf{k}} -\mu}.
\end{equation}
Here, the derivatives act on all terms to the right and $\mathbf{v} \left(\mathbf{E}\cdot\partial_{\mathbf{k}}\right) \to \left\langle \hat{\mathbf{j}}_{\sigma} \right\rangle \left(\mathbf{E}\cdot\partial_{\mathbf{k}}\right)$ in the spin current. The integrands in second-order electric and spin currents have the opposite parity compared to their counterparts in the first-order response, cf. Eq.~(\ref{8-kin-j-1-el}). Therefore, there is a spin current response to the second order in the electric field. The corresponding second-order spin response tensor $\chi_{ijl}^{(\sigma)}$ is defined as $j_{\sigma,i}^{(2)} = \chi_{ijl}^{(\sigma)} E_jE_l$ and is shown in Fig.~4 in the main text.

\end{document}


\title{Supplemental Material\\
\mbox{Minimal models and transport properties of unconventional $p$-wave magnets}}

\author{Bj{\o}rnulf Brekke}
\thanks{These authors contributed equally to this work}
\author{Pavlo Sukhachov}
\thanks{These authors contributed equally to this work}
\affiliation{Center for Quantum Spintronics, Department of Physics, Norwegian \\ University of Science and Technology, NO-7491 Trondheim, Norway}
\author{Hans Gl{\o}ckner Giil}
\affiliation{Center for Quantum Spintronics, Department of Physics, Norwegian \\ University of Science and Technology, NO-7491 Trondheim, Norway}
\author{Arne Brataas}
\affiliation{Center for Quantum Spintronics, Department of Physics, Norwegian \\ University of Science and Technology, NO-7491 Trondheim, Norway}
\author{Jacob Linder}
\affiliation{Center for Quantum Spintronics, Department of Physics, Norwegian \\ University of Science and Technology, NO-7491 Trondheim, Norway}

\maketitle
\tableofcontents

\section{Effective model}
In this section, we provide additional details of our lattice and effective models in Eqs.~(1) and (2) in the main text.

\subsection{Minimal lattice model}
\label{sec:App-1-lattice}

The lattice tight-binding model in Eq. (1) and Fig.~1a in the main text can be diagonalized in the reciprocal space. We Fourier transform the electron operators and find
\begin{align}
    H_e = \sum_{k, \ell, \ell', \sigma, \sigma'}c_{\ell,k,\sigma}^\dagger
    \big(\mathcal{H}_{k}\big)_{\ell, \ell', \sigma, \sigma'}c_{\ell',k,\sigma'},
\end{align}
where $\ell$ and $\ell'$ run over the four sites and in the magnetic unit cell, $\sigma$ and $\sigma'$ run over the spin degree of freedom and $k$ runs over the first Brillouin zone. The matrix form of the Hamiltonian is
\begin{align}
    \mathcal{H}_{\bm{k}} = \begin{pmatrix}  -\mu && J_{\mathrm{sd}} && -t e^{ik \delta } && 0 && 0 &&  0 && -t^* e^{-ik \delta } && 0 \\ J_{\mathrm{sd}} && -\mu &&
      0 && -t e^{ik \delta } && 0 && 0 && 0 && -t^*e^{-ik \delta } \\ -t^*e^{-ik \delta} && 0 && -\mu && -iJ_{\mathrm{sd}} && -t e^{ik \delta } && 0 && 0 &&  0 \\ 0 && -t^* e^{-ik \delta } && i J_{\mathrm{sd}} && -\mu &&  0 &&
      -t e^{ik \delta } &&  0 && 0 \\ 0 && 0 && -t^* e^{-ik \delta} && 0 && -\mu && - J_{\mathrm{sd}} &&  -t e^{ik \delta } && 0 \\ 0 && 0 && 0 &&
      -t^* e^{-ik \delta } && - J_{\mathrm{sd}} && -\mu && 0 && -t e^{ik \delta } \\  -t e^{ik \delta } && 0 && 0 && 0 && -t^* e^{-ik \delta } && 0 && -\mu &&  iJ_{\mathrm{sd}} \\ 0 &&
      -t e^{ik \delta } && 0 && 0 && 0 && -t^* e^{-ik \delta } && -iJ_{\mathrm{sd}} && -\mu \end{pmatrix},
    \label{3x3Matrix}
\end{align}
where $J_{\mathrm{sd}}$ is the coupling between localized and itinerant electron spins, $t$ is the nearest-neighbor hopping parameter, $\delta$ is the nearest-neighbor distance, and $\mu$ is the chemical potential.

\subsection{Effective model}
\label{sec:App-1-effective}

The effective Hamiltonian in the main text, Eq. (2), has a real space interpretation based on the main text's effective lattice in Fig.~2a. It can be interpreted as real space hopping within two sectors. It is block diagonal in terms of the two sectors
\begin{align}
     H_{\mathrm{eff}} = \begin{pmatrix} H_{\mathrm{sector}_1} && 0 \\ 0 && H_{\mathrm{sector}_2} \end{pmatrix},
    \label{App-1-effective-H}
\end{align}
where
\begin{align}
    H_{\mathrm{sector}_1} = \begin{pmatrix} \sum_{i \in (x,y)} \left[-2t \cos{(k_i a)} +\alpha_i \sin{(k_i a)}\right] - \mu  && J_{\mathrm{sd}} && \\ J_{\mathrm{sd}} &&  \sum_{i \in (x,y)} \left[-2t \cos{(k_i a)} - \alpha_i \sin{(k_i a)}\right]  - \mu
    \end{pmatrix}
    \label{sector1Matrix}
\end{align}
and
\begin{align}
    H_{\mathrm{sector}_2} = \begin{pmatrix} \sum_{i \in (x,y)} \left[-2t \cos{(k_i a)} +\alpha_i \sin{(k_i a)}\right]  - \mu && -J_{\mathrm{sd}} && \\ -J_{\mathrm{sd}} &&  \sum_{i \in (x,y)} \left[-2t \cos{(k_i a)} - \alpha_i \sin{(k_i a)} \right]- \mu
    \end{pmatrix}.
    \label{sector2Matrix}
\end{align}

The dispersion relation and the spin polarization in the model (\ref{App-1-effective-H}) are given in Eqs.~(3) and (4) in the main text, which, for convenience, we reproduce below
\begin{equation}
\label{App-1-epsilon}
\varepsilon_{\pm} = -2t \left[\cos{(k_x a)} + \cos{(k_y a)}\right]
\pm \sqrt{J_{\mathrm{sd}}^2 + \big[\alpha_x\sin{(k_x a)} + \alpha_y\sin{(k_y a)}\big]^2}
\end{equation}
and
\begin{equation}
\label{App-1-Sz}
\left<S_z\right> = \pm 2\frac{\alpha_x \sin{(k_x a)} +\alpha_y \sin{(k_y a)}}{\sqrt{J_{\mathrm{sd}}^2 + \left[\alpha_x\sin{(k_x a)}+\alpha_y\sin{(k_y a)}\right]^2}}.
\end{equation}
The spin polarization (\ref{App-1-Sz}) is generalized to the arbitrary direction of $\bm{\alpha}$.

While each sector has a nontrivial spin texture with nonzero spin polarization along different axes, the polarization components $\left<S_x\right>$ and $\left<S_y\right>$ vanish after the summation over both sectors.

\begin{figure*}[!ht]
\centering
\begin{subfigure}[b]{\textwidth}
         \includegraphics[width=\textwidth]{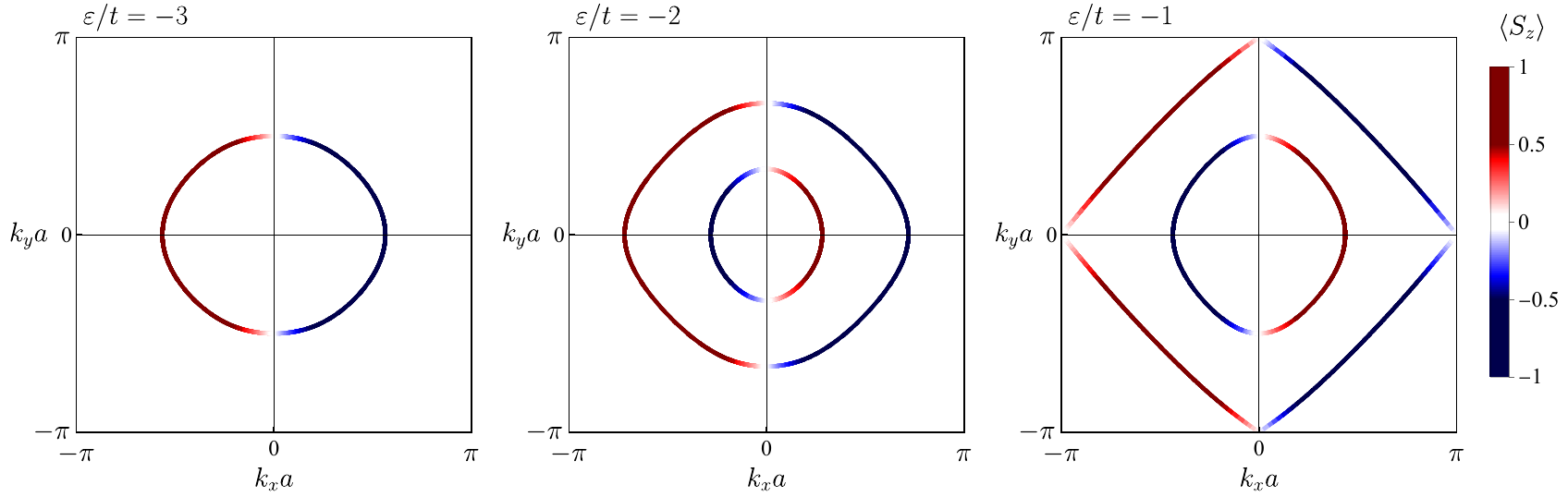}
\end{subfigure}
\begin{subfigure}[b]{\textwidth}\includegraphics[width=\textwidth]{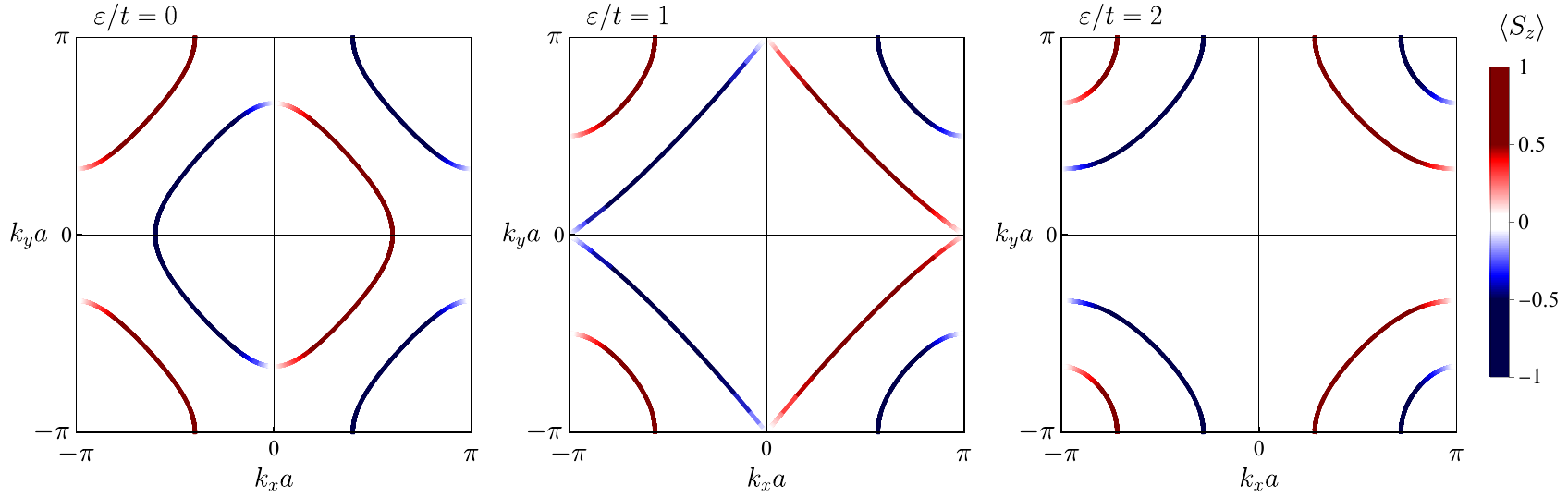}\end{subfigure}
\caption{
The constant energy contours at different values of energy $\varepsilon$. The spin polarization $\left\langle S_z \right\rangle$ is shown in red and blue colors. We use Eqs.~(\ref{App-1-epsilon}) and (\ref{App-1-Sz}) for the energy spectrum and the spin polarization, respectively. In all panels, we fix $J_{\rm sd}/t=\alpha/t=1$.
}
\label{fig:App-epsilon}
\end{figure*}

We present the constant energy contours with the spin polarization in Fig.~\ref{fig:App-epsilon}.

\subsection{Inter-sectoral hopping}
\label{sec:App-1-inter}

\begin{figure}[ht!]
\centering
\begin{subfloat}{\includegraphics[width=\textwidth]{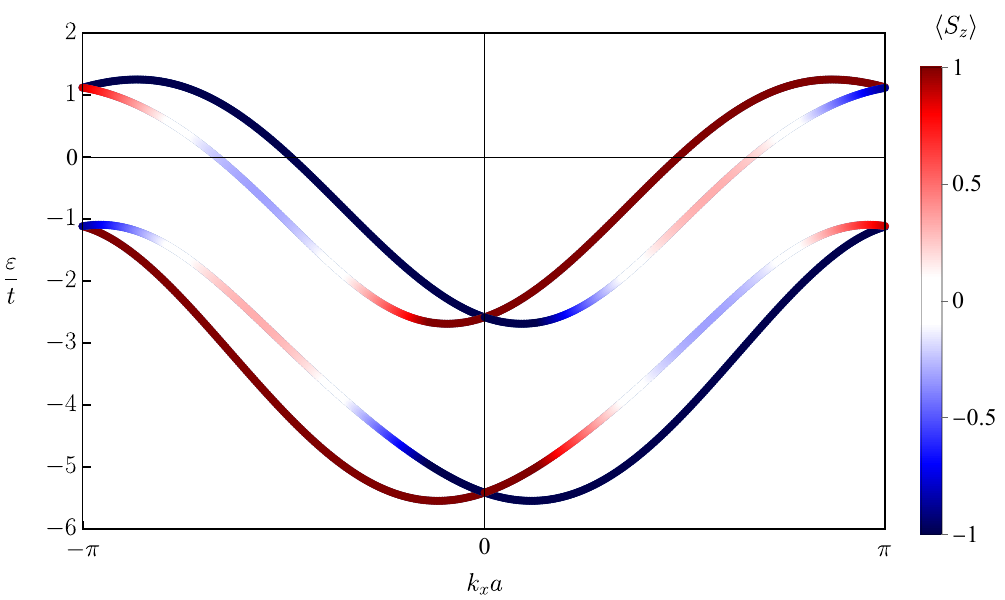}}\end{subfloat}
\begin{subfloat}{\includegraphics[width=\textwidth]{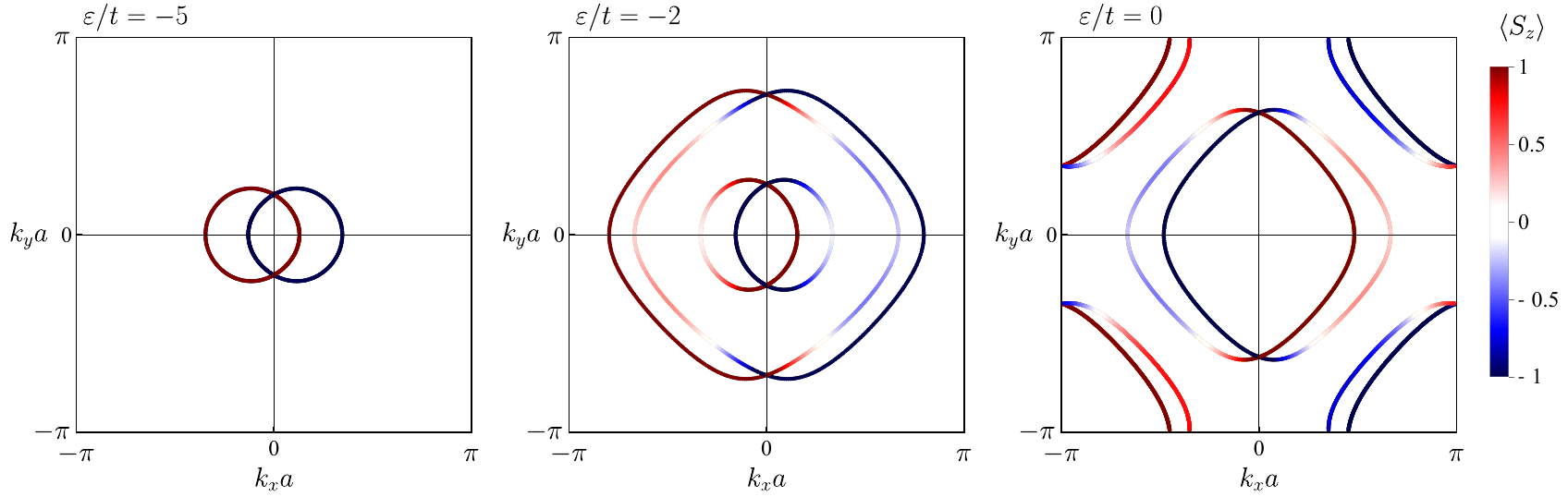}}\end{subfloat}
\caption{
The constant energy contours at different values of energy $\varepsilon$ in the presence of an inter-sectoral hopping $t_{\mathrm{inter}}/t=0.5$. The spin polarization $\left\langle S_z \right\rangle$ in units of $\hbar/2$ is shown in red and blue colors. In all panels, we fix $J_{\rm sd}/t=\alpha/t=1$.
}
\label{fig:effectiveModelIntersector-CEC}
\end{figure}

In the previous section and the main text, we consider an effective model without inter-sectoral coupling. In doing this, we retain a block diagonal form of the Hamiltonian while still capturing the physics of unconventional $p$-wave magnets. In this section, we further explore the effect of an inter-sectoral coupling in the form of a spin-independent hopping term $t_{\mathrm{inter}}$. In the presence of the inter-sectoral hopping, the effective Hamiltonian is
\begin{align}
    \nonumber
    H_{\mathrm{eff}} =& -\left\{2t \left[\cos{(k_x a)} + \cos{(k_y a)}\right] +\mu \right\}\sigma_0 \otimes \tau_0 \\ \nonumber
    &+ \left[\alpha_x \sin{(k_x a)} + \alpha_y \sin{(k_y a)}\right] \sigma_{z'} \otimes \tau_0 \\ &+ J_{\mathrm{sd}} \sigma_{x'} \otimes \tau_z + t_{\mathrm{inter}} \left[\cos{(k_x a/2)} +\cos{(k_y a/2)}\right] \sigma_0 \otimes \tau_x.
    \label{effectiveModelIntersector}
\end{align}
The last term slightly alters the electronic spectrum by inducing a splitting of the two previously degenerate bands while preserving the $p$-wave spin polarization. This split electronic spectrum is shown in Fig. \ref{fig:effectiveModelIntersector-CEC}.

The constant energy contours in lattice and continuum models at a few fixed energies are shown in Fig.~\ref{fig:effectiveModelIntersector-CEC}.

\subsection{Topology}
\label{sec:App-1-topology}

The proposed model can be generalized to include nontrivial topology. In the case of the effective model, momentum-dependent sd-coupling is sufficient.

Let us illustrate this in a 1D version of our model. For example, modifying $J_{\rm sd} \to J_{1} \cos{(k_xa)}$ in Eqs.~(\ref{App-1-effective-H}), (\ref{sector1Matrix}), and (\ref{sector2Matrix}) allows for a nontrivial winding number per sector
\begin{equation}
\label{4-model-top-nu-fin}
\nu_{\eta} = \frac{1}{2\pi} \int d k_x \psi^{\dag}_{\eta}i \sigma_y \partial_{k_x} \psi_{\eta}
= -\frac{\eta}{2\pi} 
\int d k_x\frac{J_{1}\alpha}{J_{1}^2\cos^2{(k_xa)} + \alpha^2 \sin^2{(k_xa)}},
\end{equation}
where $\psi_{\eta}$ is the eigenstate for each of the sectors $\eta=\pm$. The winding number depends nontrivially on $J_{1}$ and $\alpha$:
\begin{equation}
\label{4-model-top-nu-1}
\nu_{\eta} = \begin{cases}
    -1 & \sqrt{\left|\frac{\eta J_{1}-\alpha}{\eta J_{1}+\alpha}\right|}<1\\
    1 & \sqrt{\left|\frac{\eta J_{1}-\alpha}{\eta J_{1}+\alpha}\right|} \geq 1
    \end{cases}.
\end{equation}

The nontrivial topology is also present for a model with both exchange interactions and spin-dependent hopping. In this case, one should replace $J_{\rm sd} \to J_{\rm sd} +J_1 \cos{(k_x a)}$. The phase diagram of the system, however, becomes more complicated and is determined by $J_{\rm sd}$, $J_1$, and $\alpha$.

Let us consider a domain wall at $x=0$ with $\alpha>0$ at $x>0$ and $\alpha<0$ at $x<0$; the coefficient $J_{1}$ is assumed to be the same. Then, according to Eq.~(\ref{4-model-top-nu-1}), the winding numbers $\nu_{\eta}$ are different at $x>0$ and $x<0$. Due to the bulk-boundary correspondence, the difference between winding numbers signifies the presence of two (per sector) bound states at the interface. If sectors do not intermix, there should be four topological states. 

\section{Tunneling magnetoresistance (TMR)}
This section calculates the tunneling magnetoresistance (TMR) in unconventional $p$-wave magnets. We focus on two types of junctions composed of 2D $p$-wave magnets: (i) bilayer and (ii) planar junctions.

To make an analytical advance, we linearize the effective model in Eqs.~(\ref{App-1-effective-H})--(\ref{sector2Matrix}) around the $\Gamma$ point:
\begin{equation}
\label{2-tmr-H-i-lin}
H_{\eta}^{\rm (lin)}(\mathbf{k}) =
\left(\frac{k^2}{2m} -\mu\right) 
\sigma_0 + \left(\bm{\alpha}'\cdot\mathbf{k}\right)\sigma_{z} + \eta J_{\rm sd} \sigma_x,
\end{equation}
where $1/m=2ta^2$, $\bm{\alpha}' = a\bm{\alpha}$, and $\eta=\pm$; to simplify the notations, we will omit the prime and use $\bm{\alpha}$ instead of $\bm{\alpha}'$. Equation (\ref{2-tmr-H-i-lin}) is given as Eq.~(5) in the main text.

We find it convenient to use dimensionless variables:
\begin{equation}
\label{2-tmr-dimless}
\tilde{J} = \frac{J_{\rm sd}}{\mu}, \quad \quad \tilde{\bm{\alpha}} = \frac{2m \bm{\alpha}}{k_F}, \quad \quad \tilde{k} = \frac{k}{k_F}
\end{equation}
with $k_F=\sqrt{2m\mu}$.

The tunneling current between the $p$-wave magnets reads~\cite{Levitov-Shytov:book,Mahan:book-2013}:
\begin{equation}
\label{2-tmr-I}
I(V) = 4e\pi^3 \sum_{\eta=\pm} \int_{-\infty}^{\infty} d\omega \sum_{\mathbf{k},\mathbf{p}} |T_{\mathbf{p},\mathbf{k}}|^2  \mbox{Tr}{\left\{\IM{G_{1;\eta}(\omega;\mathbf{p})} \IM{G_{2;\eta}(\omega - eV;\mathbf{k})}\right\}} \left[f_{2}(\omega-eV) -f_{1}(\omega)\right].
\end{equation}
Here, $G_{i;\eta}(\omega;\mathbf{p})$ is the retarded Green's function of the $i$-th magnet, $T_{\mathbf{p},\mathbf{k}}$ is the tunneling coefficient, $f_{i}(\omega)$ is the Fermi-Dirac distribution function, and $V$ is the voltage bias between the magnets.   

The tunneling coefficient for a bilayer preserves in-plane momenta: $|T_{\mathbf{p},\mathbf{k}}|^2 = T_0^2 \delta_{\mathbf{p},\mathbf{k}}$. In the case of planar interfaces, only the momentum components along the interface are conserved: $|T_{\mathbf{p},\mathbf{k}}|^2 = T_0^2 \delta_{p_{\parallel},k_{\parallel}}/L$ with $L$ being the size of the interface.

The Green's function is defined as 
\begin{equation}
\label{2-tmr-Green-lin}
G_{i;\eta}(\omega;\mathbf{k}) = \frac{1}{\omega +i0^{+} - H_{\eta}^{\rm (lin)}(\mathbf{k})}
=\frac{\left(\tilde{\omega}+\tilde{k}^2-1\right) \sigma_0 + \eta \tilde{J}_i \sigma_x +\left(\tilde{\bm{\alpha}}_i\cdot\tilde{\mathbf{k}}\right) \sigma_z}{\left(\tilde{\omega} +i0^{+} +\tilde{k}^2-1\right)^2 -\tilde{J}_i^2 -\left(\tilde{\bm{\alpha}}_i\cdot\tilde{\mathbf{k}}\right)^2},
\end{equation}
where we used the Hamiltonian (\ref{2-tmr-H-i-lin}) in the second expression.

The imaginary part is straightforwardly obtained by using the Sokhotski–Plemelj theorem,
\begin{eqnarray}
\label{2-tmr-Green-lin-Im}
\IM{G_{i;\eta}(\omega;\mathbf{k})} &=& -\pi \sign{\tilde{\omega}+ \tilde{k}^2-1} 
\df{\left(\tilde{\omega} +\tilde{k}^2-1\right)^2 -\tilde{J}_i^2 -\left(\tilde{\bm{\alpha}}_i\cdot\tilde{\mathbf{k}}\right)^2 } \nonumber\\
&\times& \left[\left(\tilde{\omega}+\tilde{k}^2-1\right) \sigma_0 + \eta \tilde{J}_i \sigma_x +\left(\tilde{\bm{\alpha}}_i\cdot\tilde{\mathbf{k}}\right) \sigma_z\right].
\end{eqnarray}

We focus on the differential conductance $G=dI/dV$, assume the low-temperature limit $T\to0$, and consider the limit of small voltage biases $|eV|\ll \mu$. This allows us to simplify the expression for the differential conductance:
\begin{equation}
\label{2-tmr-G}
G = \frac{dI}{dV} = 4e\pi^3 \sum_{\eta=\pm} \sum_{\mathbf{k},\mathbf{p}} |T_{\mathbf{p},\mathbf{k}}|^2  \mbox{Tr}{\left\{\IM{G_{1;\eta}(0;\mathbf{p})} \IM{G_{2;\eta}(0;\mathbf{k})}\right\}}.
\end{equation}

\subsection{Bilayer}
\label{sec:2-tmr-i-bilayer}

In the bilayer, the differential conductance (\ref{2-tmr-G}) is rewritten as
\begin{eqnarray}
\label{2-tmr-G-bilayer}
G(\theta_{\alpha}) &=& e\pi^2 k_F^2 T_0^2 \int_0^{2\pi}\frac{d\theta}{2\pi} 
\sum_{\tilde{k}_{\pm}}\df{\left(\tilde{k}_{\pm}^2-1\right)^2 -\tilde{J}_2^2 -\tilde{\alpha}_2^2\tilde{k}_{\pm}^2\cos^2{\left(\frac{\theta-\theta_{\alpha}}{2}\right)}} \nonumber\\
&\times&\left[\left(\tilde{k}_{\pm}^2-1\right)^2 +\tilde{J}_1\tilde{J}_2 +\tilde{k}_{\pm}^2 \frac{\tilde{\alpha}_1\tilde{\alpha}_2}{2} \left(\cos{\theta} +\cos{\theta_{\alpha}}\right)\right],
\end{eqnarray}
where we used Eq.~(\ref{2-tmr-Green-lin-Im}), defined
\begin{equation}
\label{2-tmr-bilayer-kpm}
\tilde{k}_{\pm}^2 = 1 +\frac{\tilde{\alpha}_1^2}{2} \cos^2{\left(\frac{\theta+\theta_{\alpha}}{2}\right)} \pm \sqrt{\tilde{J}_1^2 +\tilde{\alpha}_1^2 \cos^2{\left(\frac{\theta+\theta_{\alpha}}{2}\right)} \left[1 +\frac{\tilde{\alpha}_1^2}{4} \cos^2{\left(\frac{\theta+\theta_{\alpha}}{2}\right)}\right]},
\end{equation}
and introduced the angle $\theta_{\alpha}$ between the spin-splitting vectors in two layers, $\bm{\alpha}_1$ and $\bm{\alpha}_2$.

The $\delta$-function in Eq.~(\ref{2-tmr-G-bilayer}) is treated as a Lorentzian, $\delta(x)\to \delta_{\Gamma}(x)$:
\begin{equation}
\label{2-tmr-delta}
\delta_{\Gamma}(x) = \frac{1}{\pi} \frac{\Gamma}{x^2+\Gamma^2}.
\end{equation}

As in the main text, see Eq.~(6), we define the TMR as
\begin{equation}
\label{2-TMR-def}
\mathrm{TMR}= \frac{G(\theta_{\alpha})-G(\theta_{\alpha}+\pi)}{G(\theta_{\alpha}+\pi)}.
\end{equation}

\begin{figure*}[!ht]
\centering
\begin{subfigure}[b]{0.31\textwidth}
         \includegraphics[width=\textwidth]{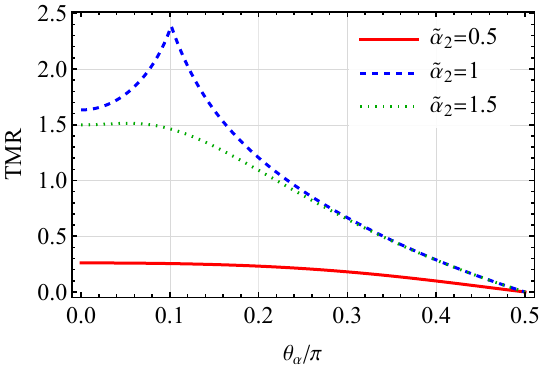}
         \caption{$\tilde{J}_2=0.5$}
\end{subfigure}
\begin{subfigure}[b]{0.31\textwidth}
         \includegraphics[width=\textwidth]{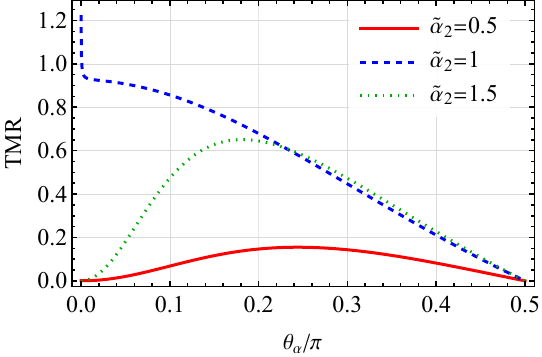}
         \caption{$\tilde{J}_2=1$}
\end{subfigure}
\begin{subfigure}[b]{0.31\textwidth}
         \includegraphics[width=\textwidth]{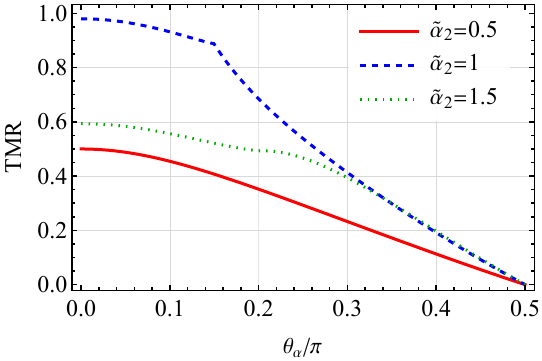}
         \caption{$\tilde{J}_2=1.5$}
\end{subfigure}
\caption{
TMR as a function of the relative angle $\theta_{\alpha}$ between the spin-splitting vectors $\bm{\alpha}_1$ and $\bm{\alpha}_2$ in two layers of the bilayer. In all panels, $\tilde{\alpha}_1=1$, $\tilde{J}_1=1$, and $\tilde{\Gamma} = 10^{-3}$.
}
\label{fig:2-tmr-i-bilayer-lin}
\end{figure*}

We present the TMR for the bilayer in Fig.~\ref{fig:2-tmr-i-bilayer-lin}. The strong dependence on $\tilde{J}_2$ shown in Fig.~\ref{fig:2-tmr-i-bilayer-lin} originates primarily from the mismatch of the Fermi surfaces rather than the spin polarization. As one can see from Figs.~\ref{fig:2-tmr-i-bilayer-lin}(a) and \ref{fig:2-tmr-i-bilayer-lin}(b), perfect matching of parameters $\tilde{J}$ may not be needed to achieve large TMR. However, matching the spin-splitting vectors is desirable.

\subsection{Planar junction}
\label{sec:2-tmr-i-planar}

In the case of the planar junction, we have the following expression for the conductance (\ref{2-tmr-G}):
\begin{eqnarray}
\label{2-tmr-i-planar-G}
G(\theta_{\alpha, 1},\theta_{\alpha, 2}) &=& e\pi^3 T_0^2 k_F^3\int_{0}^{2\pi}\frac{d\theta_1}{2\pi} \int_0^{\infty}\tilde{k} d\tilde{k} \int_{-\infty}^{\infty}d\tilde{p}_{\perp}
\sign{\tilde{k}^2-1} \sign{\tilde{k}^2\sin^2{\theta_1}+\tilde{p}_{\perp}^2-1} \nonumber\\
&\times&
\df{\left(\tilde{k}^2-1\right)^2 -\tilde{J}_1^2 -\tilde{k}^2\tilde{\alpha}_{1}^2\left(\cos{\theta_1} \cos{\theta_{\alpha,1}} +\sin{\theta_1} \sin{\theta_{\alpha,1}}\right)^2}\nonumber\\
&\times&\df{\left(\tilde{k}^2\sin^2{\theta_1}+\tilde{p}_{\perp}^2-1\right)^2 -\tilde{J}_2^2 -\tilde{\alpha}_{2}^2\left(\tilde{p}_{\perp}\cos{\theta_{\alpha,2}} +\tilde{k}\sin{\theta_1} \sin{\theta_{\alpha,2}}\right)^2} \nonumber\\
&\times&\Bigg[\tilde{J}_1\tilde{J}_2 +\left(\tilde{k}^2-1\right)\left(\tilde{k}^2\sin^2{\theta} +\tilde{p}_{\perp}^2 -1\right) \nonumber\\
&+&\tilde{k}\tilde{\alpha}_1\tilde{\alpha}_2\left(\cos{\theta_1} \cos{\theta_{\alpha,1}} +\sin{\theta_1} \sin{\theta_{\alpha,1}}\right) \left(\tilde{p}_{\perp} \cos{\theta_{\alpha,2}} +\tilde{k}\sin{\theta_1} \sin{\theta_{\alpha,2}}\right)
\Bigg],
\end{eqnarray}
where the angles are calculated with respect to the normal to the junction. The $\delta$-functions allow us to straightforwardly integrate over $\tilde{k}$ and $\tilde{p}_{\perp}$; the integral over $\theta_1$ is taken numerically. 

Since there is a preferred direction in a planar interface, the junction cannot be characterized by a single angle $\theta_{\alpha}$. Then, the definition of the TMR (\ref{2-TMR-def}) is modified as
\begin{equation}
\label{2-TMR-planar-def}
\mathrm{TMR}= \frac{G(\theta_{\alpha,1},\theta_{\alpha,2})-G(\theta_{\alpha,1},\theta_{\alpha,2}+\pi)}{G(\theta_{\alpha,1}, \theta_{\alpha,2}+\pi)}.
\end{equation}

\begin{figure*}[!ht]
\centering
\begin{subfigure}[b]{0.31\textwidth}
         \includegraphics[width=\textwidth]{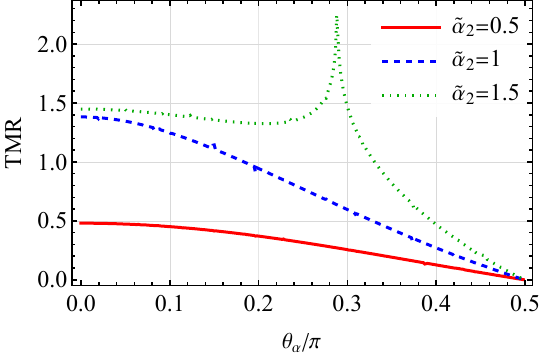}
         \caption{$\tilde{J}_2=0.5$}
\end{subfigure}
\begin{subfigure}[b]{0.31\textwidth}
         \includegraphics[width=\textwidth]{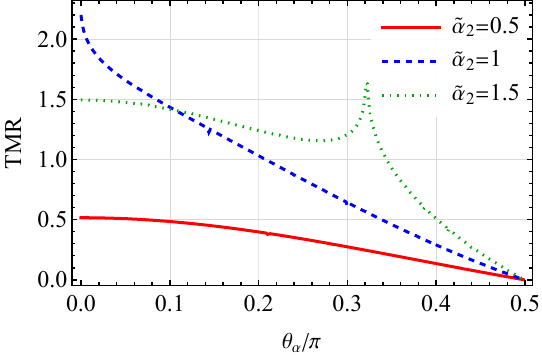}
         \caption{$\tilde{J}_2=1$}
\end{subfigure}
\begin{subfigure}[b]{0.31\textwidth}
         \includegraphics[width=\textwidth]{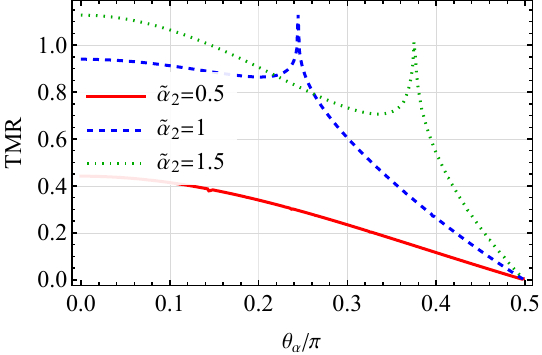}
         \caption{$\tilde{J}_2=1.5$}
\end{subfigure}
\caption{
TMR as a function of the angle $\theta_{\alpha} = \theta_{\alpha,2}-\theta_{\alpha,1}$ between the spin-splitting vectors $\bm{\alpha}_1$ and $\bm{\alpha}_2$ in two magnets of the planar junction. We fixed $\theta_{\alpha,1}=\pi/2$ in Eqs.~(\ref{2-tmr-i-planar-G}) and (\ref{2-TMR-planar-def}). In all panels, $\tilde{J}_1=1$ and $\tilde{\alpha}_1=1$.
}
\label{fig:2-tmr-i-planar-lin}
\end{figure*}

We present the TMR for the planar interface in Fig.~\ref{fig:2-tmr-i-planar-lin}. Compared to the bilayer, see Fig.~\ref{fig:2-tmr-i-bilayer-lin}, the TMR in the planar interface is less sensitive to the mismatch of model parameters and can show larger values of the TMR.

Furthermore, owing to the interplay of the spin texture and the overlap of the Fermi surfaces, there are well-pronounced peaks in the TMR. As we also discuss in the main text, see Fig.~3(c), the peak is observed when the Fermi surface in the magnet that is rotated is larger than that in the fixed magnet. In the case of equal parameters $\tilde{J}_1=\tilde{J}_2$, this requires $\tilde{\alpha}_2>\tilde{\alpha}_1$. Rotating away from the collinear orientation of the spin-splitting vectors results in two effects: (i) smaller asymmetry of the spin texture at the spin flip and (ii) change of the overlap between the Fermi surfaces. The former effect is observed for any size of the Fermi surface in the rotated magnet and is the dominant effect at $\tilde{\alpha}_2<\tilde{\alpha}_1$, see Fig.~\ref{fig:2-tmr-i-planar-lin}. On the other hand, the second effect requires the Fermi surface in the rotated magnet (right magnet) to be larger than in the fixed one (left magnet). For small rotation angles, the increase of the overlap increases the conductance of the junction allowing for a larger TMR, but this increase is mostly overcome by the symmetrization of the spin texture resulting in a slower decline. The peak is observed at the maximal overlap $\theta_{\alpha}\approx \theta_{\alpha, {\rm cr}}$. Fixing $\theta_{\alpha,1}=\pi/2$, the case of arbitrary $\tilde{J}_i$ and $\tilde{\alpha}_i$, the characteristic angle reads
\begin{equation}
\label{2-TMR-theta-cr}
\theta_{\alpha, {\rm cr}} = \arccos{\left\{\sqrt{\frac{2\left(\tilde{J}_1^2 -\tilde{J}_2^2\right) +\tilde{\alpha}_1^2 \left(2 + \tilde{\alpha}_1^2 +\sqrt{4\tilde{J}_1^2 +4\tilde{\alpha}_1^2 +\tilde{\alpha}_1^4}\right)}{\tilde{\alpha}_2^2\left(2+\tilde{\alpha}_1^2 + \sqrt{4\tilde{J}_1^2 +4\tilde{\alpha}_1^2 +\tilde{\alpha}_1^4}\right)}}\right\}}.
\end{equation}
Rotating the second magnet further, $\theta_{\alpha}>\theta_{\alpha, {\rm cr}}$, decreases the overlap and reduces the role of the spin texture leading to smaller TMR.

\subsection{Role of the inter-sectoral hopping}
\label{sec:2-tmr-i-inter}

To address the role of the inter-sectoral hopping for the TMR, we employ the continuum effective model of $p$-wave magnets whose dimensionless form reads
\begin{equation}
\tilde{H}_{\mathrm{eff}}^{(\rm lin)} = \left(\tilde{k}^2 -1\right)\sigma_0 \otimes \tau_0 + \left(\tilde{\bm{\alpha}}\cdot\tilde{\mathbf{k}}\right) \sigma_{z'} \otimes \tau_0 + \tilde{J} \sigma_{x'} \otimes \tau_z + \tilde{t}_{\rm inter} \sigma_0 \otimes \tau_x,
\label{effectiveModelIntersector-continuum-dimless}
\end{equation}
where $\tilde{t}_{\rm inter}=2t_{\rm inter}/\mu$.

The TMR is calculated via the same framework as in the previous part of the section, see Eq.~(\ref{2-tmr-G}). However, since Green's functions are no longer block-diagonal in the sectoral space, the expression for the conductance (\ref{2-tmr-G}) should be adjusted, 
\begin{equation}
\label{2-tmr-G-inter}
G = 4e\pi^3 \sum_{\mathbf{k},\mathbf{p}} |T_{\mathbf{p},\mathbf{k}}|^2  \mbox{Tr}{\left\{\IM{G_{1}(0;\mathbf{p})} \IM{G_{2}(0;\mathbf{k})}\right\}},
\end{equation}
where the trace is taken both over the spin and sectoral degrees of freedom. In what follows, we illustrate the role of the inter-sectoral hopping in the case of the planar junction.

In the calculations of the conductance, we follow the same steps as in Sec.~\ref{sec:2-tmr-i-planar}. A more complicated structure of the Hamiltonian, however, leads to cumbersome expression. Nevertheless, by extracting the imaginary part of Green's function via the Sokhotski–Plemelj theorem, integrating over the parallel component of the momentum in one of the magnets and the momentum magnitude in the resulting expression, we reduce the conductance to a single integral over the angle, which is calculated numerically. 

\begin{figure}[ht!]
\centering
{\includegraphics[width=0.5\columnwidth]{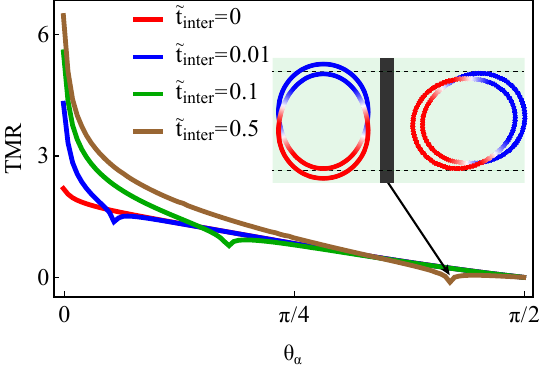}}
\caption{
TMR as a function of the relative angle $\theta_{\alpha}$ between the spin-splitting vectors $\tilde{\bm{\alpha}}_1$ and $\tilde{\bm{\alpha}}_2$ in the two magnets of the planar junction. The insets show the relative orientation of the spin-polarized Fermi surfaces. We use the same inter-sectoral coupling in both magnets $\tilde{t}_{\rm inter}= \tilde{t}_{\rm inter,1 }= \tilde{t}_{\rm inter,2}$. Other parameters are $\tilde{J}_1=\tilde{J}_2=\tilde{\alpha}_1=\tilde{\alpha}_2=1$.
}
\label{fig:TMR-tinter}
\end{figure}

We show the TMR for the $p$-wave magnets with equivalent parameters in Fig.~\ref{fig:TMR-tinter} as a function of the relative angle $\theta_{\alpha}=\theta_{\alpha,2}-\theta_{\alpha,1}$ between the spin-splitting vectors $\tilde{\bm{\alpha}}_1$ and $\tilde{\bm{\alpha}}_2$. As one can see, the inter-sectoral hopping results in a larger TMR at small $\theta_{\alpha}$ and a dip at a certain value of $\theta_{\alpha}$. Both of these features are related to the interplay of the Fermi surface structure and its spin polarization. The enhancement of the TMR at small $\theta_{\alpha}$ is related to larger spin polarization of the outer parts of the Fermi surfaces at $\tilde{t}_{\rm inter}\neq0$, see Fig.~\ref{fig:effectiveModelIntersector-CEC} and the inset in Fig.~\ref{fig:TMR-tinter} for a schematic illustration of the Fermi surfaces. The increase of $\theta_{\alpha}$ results in a more symmetric spin texture with respect to the spin flip in the rotated magnet, hence the decrease in the TMR. The dip in the TMR is observed when the inner parts of the Fermi surface in the fixed magnet match the outer parts in the rotated magnet, see the inset in Fig.~\ref{fig:TMR-tinter}. The TMR similar to that at $\tilde{t}_{\rm inter}=0$ is observed for larger $\theta_{\alpha}$.

\section{Spin filtering}
Let us show that $p$-wave magnets allow for spin filtering where the differential conductances for spin-up and spin-down particles differ. In addition, we contrast the results for our model of unconventional $p$-wave magnets with those for fully spin-polarized Fermi surfaces in Sec.~\ref{sec:8-ii}.

\subsection{Unconventional $p$-wave magnet}
\label{sec:8-i}

We consider a planar interface between an unconventional $p$-wave magnet and a regular metal. The Hamiltonian for the interface is
\begin{equation}
\label{8-i-H-lin}
H_{\eta}(x,k_y) =\left(-\frac{\nabla^2_{x}}{2m} +\frac{k_{\parallel}^2}{2m} -\mu\right) \sigma_0 + \left[-\frac{i}{2}\left\{\alpha_x(x),\nabla_x\right\} +\alpha_y(x) k_{y}\right]\sigma_{z} + \eta J_{\rm sd}(x) \sigma_x +U(x)\sigma_0,
\end{equation}
where $\bm{\alpha}(x) = \bm{\alpha} \Theta{(x)}$ and $J_{\rm sd}(x) = J_{\rm sd} \Theta{(x)}$ with $\Theta{(x)}$ being the unit-step function; the barrier at the interface is modeled as $U(x)=U\delta(x)$~\cite{BTK:1982}. For definiteness, we fixed $x$- and $y$-axes to be perpendicular and parallel to the interface, respectively. Note that the form of the term $\left\{\alpha_x(x),\nabla_x\right\}$ with $\left\{\ldots,\ldots\right\}$ being the anticommutator is enforced by the hermiticity of the Hamiltonian.

We determine the differential conductance in the metallic part of the junction assuming impinging electrons with the spin projection $s$ via the standard formulas:
\begin{equation}
\label{8-i-G-e-0}
G_{e;s}(V) = -e^2 L^3 \sum_{\eta=\pm}\int \frac{d\mathbf{k}}{(2\pi)^2} j_{e;\eta,s} f'(\varepsilon-eV)
=-e^2 L^3 \sum_{\eta=\pm}\int \frac{dk_y}{(2\pi)^2} \int d\varepsilon \frac{\partial k_x}{\partial \varepsilon} j_{e;\eta,s} f'(\varepsilon-eV)
\end{equation}
for the electric conductance and
\begin{equation}
\label{8-i-G-s-0}
G_{\sigma;s}(V) = -e^2 L^3 \sum_{\eta=\pm}\int \frac{dk_y}{(2\pi)^2} \int d\varepsilon \frac{\partial k_x}{\partial \varepsilon} j_{\sigma;\eta,s} f'(\varepsilon-eV)
\end{equation}
for the spin conductance. Here, prime denotes the derivative with respect to the argument. The summation over the spin projections should be performed in the full electric and spin conductances. The electric and spin currents used in Eqs.~(\ref{8-i-G-e-0}) and (\ref{8-i-G-s-0}) are
\begin{eqnarray}
\label{8-i-je-def}
j_{e;\eta,s} &=& \frac{1}{2L^2}\RE{\psi_{\eta,s}^{\dag} \overleftrightarrow{\hat{v}}_x \psi_{\eta,s}},\\
\label{8-i-js-def}
j_{\sigma;\eta,s} &=& \frac{1}{2L^2}\RE{\psi_{\eta,s}^{\dag} \sigma_z\overleftrightarrow{\hat{v}}_x \psi_{\eta,s}},
\end{eqnarray}
respectively. For the current in the metallic part of the junction, $\hat{v}_x = -i\nabla_{x}/m$, and $\psi_{\eta,s}$ denote the scattering states.
The scattering states for impinging spin-up and spin-down particles are:
\begin{eqnarray}
\label{8-i-psi-s-up-1}
\psi_{\eta,+}(x<0) &=& e^{ik_xx +ik_yy} \begin{pmatrix} 
1\\
0
\end{pmatrix}
+r_{1,\eta,+} e^{-ik_xx -ik_yy} \begin{pmatrix} 
1\\
0
\end{pmatrix} 
+r_{2,\eta,+} e^{-ik_xx -ik_yy} \begin{pmatrix} 
0\\
1
\end{pmatrix},\\
\label{8-i-psi-s-up-2}
\psi_{\eta,+}(x>0) &=& t_{1,\eta,+} e^{iq_{x,+,+}x +ik_yy} \psi_{\eta}(q_{x,+,+}) + t_{2,\eta,+} e^{iq_{x,+,-}x +ik_yy} \psi_{\eta}(q_{x,+,-})
\end{eqnarray}
and
\begin{eqnarray}
\label{8-i-psi-s-down-1}
\psi_{\eta,-}(x<0) &=& e^{ik_xx +ik_yy} \begin{pmatrix} 
0\\
1
\end{pmatrix}
+r_{1,\eta,-} e^{-ik_xx -ik_yy} \begin{pmatrix} 
1\\
0
\end{pmatrix} 
+r_{2,\eta,-} e^{-ik_xx -ik_yy} \begin{pmatrix} 
0\\
1
\end{pmatrix},\\
\label{8-i-psi-s-down-2}
\psi_{\eta,-}(x>0) &=& t_{1,\eta,-} e^{iq_{x,+,+}x +ik_yy} \psi_{\eta}(q_{x,+,+}) + t_{2,\eta,-} e^{iq_{x,+,-}x +ik_yy} \psi_{\eta}(q_{x,+,-}),
\end{eqnarray}
respectively. Here, $k_x = k_F\sqrt{\tilde{\varepsilon} -\tilde{k}_y^2}$,
\begin{equation}
\label{8-i-psi}
\psi_{\eta}(k_x) =
\frac{\left|\tilde{\varepsilon} -\tilde{k}^2 +\left(\tilde{\bm{\alpha}}\cdot \tilde{\mathbf{k}}\right)\right|}{\sqrt{\left[\tilde{\varepsilon} -\tilde{k}^2 +\left(\tilde{\bm{\alpha}}\cdot \tilde{\mathbf{k}}\right)\right]^2 +\tilde{J}^2}} \begin{pmatrix}
1\\
\frac{\eta \tilde{J}}{\tilde{\varepsilon} -\tilde{k}^2 +\left(\tilde{\bm{\alpha}}\cdot \tilde{\mathbf{k}}\right)}
\end{pmatrix},
\end{equation}
and $q_{x,s_1,s_2}= k_F \tilde{q}_{x,s_1,s_2}$ with
\begin{eqnarray}
\label{8-i-qx-alphax0}
\tilde{\alpha}_y=0: &\quad& \tilde{q}_{x,s_1,s_2} = s_1 \sqrt{\tilde{\varepsilon}  +\frac{\tilde{\alpha}_x^2}{2} -\tilde{k}_y^2 +s_2 \sqrt{\tilde{J}^2 +\tilde{\alpha}_x^2\left(\frac{\tilde{\alpha}_x^2}{4} +\tilde{\varepsilon}  -\tilde{k}_y^2 \right)}},\nonumber\\
\\
\label{8-i-qx-alphay0}
\tilde{\alpha}_x=0: &\quad& \tilde{q}_{x,s_1,s_2} = s_1 \sqrt{\tilde{\varepsilon}  -\tilde{k}_y^2 +s_2\sqrt{\tilde{J}^2 +\tilde{\alpha}_y^2 \tilde{k}_y^2}}.
\end{eqnarray}
The wave vectors $\tilde{q}_{x,s_1,s_2}$ are obtained from
\begin{equation}
\label{8-i-eps}
\tilde{\varepsilon} = \tilde{q}^2 \pm \sqrt{\tilde{J}^2 + \left(\tilde{\bm{\alpha}} \cdot \tilde{\mathbf{q}}\right)^2},
\end{equation}
albeit are cumbersome for a generic orientation of $\tilde{\bm{\alpha}}$.

Note that since the Hamiltonian (\ref{8-i-H-lin}) is not diagonal in the spin space, we include the reflected waves of different spin polarizations and introduce two reflection coefficients: one ($r_{1,\eta,+}$ or $r_{2,\eta,-}$) corresponds to the reflection without the spin-flip and the other ($r_{2,\eta,+}$ or $r_{1,\eta,-}$) describes the spin-flip process.

The boundary conditions for the planar interface are
\begin{eqnarray}
\label{8-i-BC-1}
x=0: &\quad& \psi_{\eta,s}(x>0)=\psi_{\eta,s}(x<0),\\
\label{8-i-BC-2}
x=0: &\quad& \partial_x\psi_{\eta,s}(x>0)-\partial_x\psi_{\eta,s}(x<0)=\left(Z -i\frac{\tilde{\alpha}_x}{2} \sigma_z\right) \psi_{\eta,s}(x>0),
\end{eqnarray}
where the dimensionless parameter $Z=2mU/k_F$ quantifies the strength of the potential barrier between the metal and the $p$-wave magnet. The last term in the parentheses in Eq.~(\ref{8-i-BC-2}) originates from the term $-\frac{i}{2}\left\{\alpha_x \Theta{(x)},\nabla_x\right\}$ in Eq.~(\ref{8-i-H-lin}).

Substituting Eqs.~(\ref{8-i-psi-s-up-1}), (\ref{8-i-psi-s-up-2}), (\ref{8-i-psi-s-down-1}), and (\ref{8-i-psi-s-down-2}) into Eqs.~(\ref{8-i-je-def}) and (\ref{8-i-js-def}), as well as assuming the zero-temperature limit, we obtain the following electric (\ref{8-i-G-e-0}) and spin (\ref{8-i-G-s-0}) conductances:
\begin{eqnarray}
\label{8-i-G-e}
G_e(V) &=& e^2 L \sum_{\eta,s=\pm}\int \frac{dk_y}{(2\pi)^2} \int d\varepsilon \left(1-|r_{1,s}|^2-|r_{2,s}|^2\right) \df{\varepsilon -\mu -eV} \nonumber\\
&=& \frac{\tilde{G}_{Q}}{2}\sum_{\eta,s=\pm}\int d\tilde{k}_y \left(1-|r_{1,s}|^2-|r_{2,s}|^2\right)\Big|_{\tilde{\varepsilon} = eV/\mu+1}
\end{eqnarray}
and
\begin{eqnarray}
\label{8-i-G-s}
G_{\sigma}(V) &=& e^2 L \sum_{\eta,s=\pm}\int \frac{dk_y}{(2\pi)^2} \int d\varepsilon \left(s-|r_{1,s}|^2+|r_{2,s}|^2\right) \df{\varepsilon -\mu -eV} \nonumber\\
&=& \frac{\tilde{G}_{Q}}{2}\sum_{\eta,s=\pm}\int d\tilde{k}_y \left(s-|r_{1,s}|^2+|r_{2,s}|^2\right)\Big|_{\tilde{\varepsilon} = eV/\mu+1}.
\end{eqnarray}
Here, $\tilde{G}_{Q}=G_Q k_F L/(2\pi)$ with $G_{Q}=e^2/\pi$. To quantify the spin filtering, we introduce the following conductance asymmetry:
\begin{equation}
\label{8-i-Delta-G-def}
\Delta G(V) = \frac{G_{e,\downarrow}(V)-G_{e,\uparrow}(V)}{G_{e,\uparrow}(V)+G_{e,\downarrow}(V)},
\end{equation}
where
\begin{equation}
\label{8-i-G-e-s}
G_{e,s}(V) = \frac{\tilde{G}_{Q}}{2}\sum_{\eta=\pm}\int d\tilde{k}_y \left(1-|r_{1,s}|^2-|r_{2,s}|^2\right)\Big|_{\tilde{\varepsilon} = eV/\mu+1}
\end{equation}
is the electric conductance per impinging spin.

\begin{figure*}[!ht]
\centering
\begin{subfigure}[b]{0.31\textwidth}
         \includegraphics[width=\textwidth]{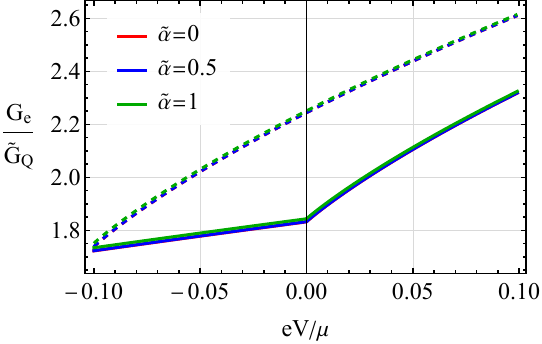}
\end{subfigure}
\begin{subfigure}[b]{0.31\textwidth}
         \includegraphics[width=\textwidth]{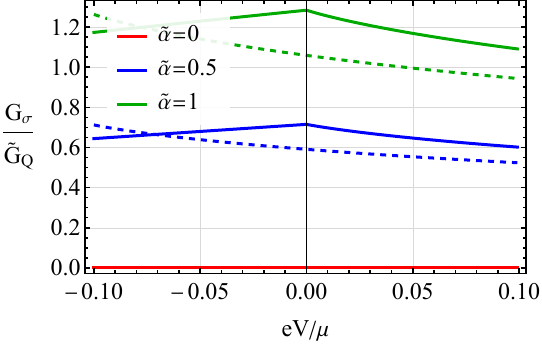}
\end{subfigure}
\begin{subfigure}[b]{0.31\textwidth}
         \includegraphics[width=\textwidth]{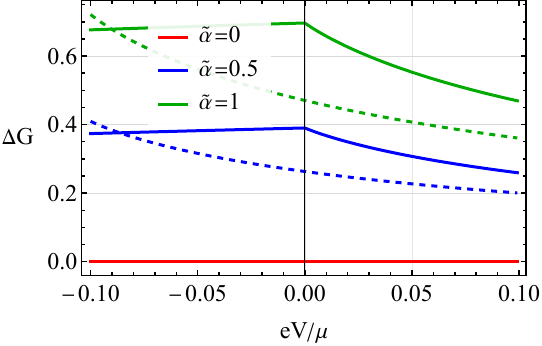}
\end{subfigure}
\caption{
The electric $G_e$ (panel (a)) and spin $G_{\sigma}$ (panel (b)) conductances as well as the conductance asymmetry $\Delta G$ (panel (c)). Solid and dashed lines correspond to $\tilde{J}=1$ and $\tilde{J}=0.9$, respectively.
The spin-splitting vector $\tilde{\bm{\alpha}}$ is perpendicular to the interface, $\tilde{\alpha}_y=0$.
We use Eqs.~(\ref{8-i-G-e}), (\ref{8-i-G-s}), and (\ref{8-i-Delta-G-def}). In all panels, we set $Z=0$.
}
\label{fig:8-i-J=1-alphay=0}
\end{figure*}

We show the electric and spin conductance as well as the anisotropy (\ref{8-i-Delta-G-def}) for the spin-splitting vector perpendicular to the interface ($\tilde{\alpha}_y=0$) in Fig.~\ref{fig:8-i-J=1-alphay=0} at $\tilde{J}=1$ (solid lines) and $\tilde{J}=0.9$ (dashed lines), respectively. As one can see, the main effect of a smaller $\tilde{J}$ is to reduce the voltage onset needed for a nontrivial conductance. The onset, i.e., the value of the voltage bias that should be reached for a nonzero conductance, is determined by requiring at least one of the wave vectors of electrons in the magnet to be real $\RE{\tilde{q}_{x,+,+}}\geq0$, i.e., there should be propagating electrons in the magnet; for propagating waves in the metal, the wave vector $\tilde{k}_x$ should be real as well. To estimate the threshold, it is sufficient to consider the shortest trajectories and set $\tilde{k}_y=0$ in $\tilde{k}_x$ leading to $eV_{\rm cr} = -\mu$. At $eV> |J_{\rm sd}|-\mu$, the second band corresponding to the sign $+$ in Eq.~(\ref{8-i-eps}) is activated, leading to a kink-like feature in the conductance. Note that $\tilde{q}_{x,s_1,s_2}$ can be found exactly at $\tilde{k}_y=0$: it is given by Eq.~(\ref{8-i-qx-alphax0}) at $\tilde{k}_y=0$ and does not depend on $\tilde{\alpha}_y$.

In contrast to the spin conductance, shown in Fig.~\ref{fig:8-i-J=1-alphay=0}(b), which is determined by the spin-splitting vector $\tilde{\bm{\alpha}}$, the electric conductance shown in Fig.~\ref{fig:8-i-J=1-alphay=0}(a) demonstrates negligible dependence on $\tilde{\bm{\alpha}}$. The conductance asymmetry (\ref{8-i-Delta-G-def}) and, as a result, the spin filtering is maximal near the threshold and gradually decays with the bias $eV$. 

\begin{figure*}[!ht]
\centering
\begin{subfigure}[b]{0.31\textwidth}
         \includegraphics[width=\textwidth]{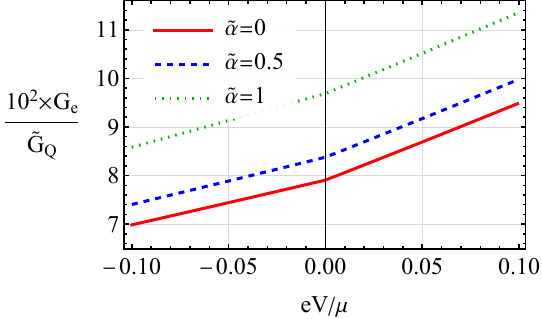}
\end{subfigure}
\begin{subfigure}[b]{0.31\textwidth}
         \includegraphics[width=\textwidth]{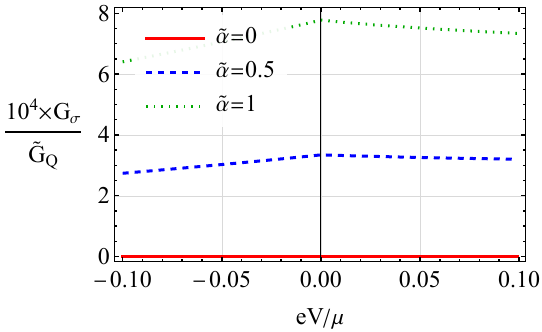}
\end{subfigure}
\begin{subfigure}[b]{0.31\textwidth}
         \includegraphics[width=\textwidth]{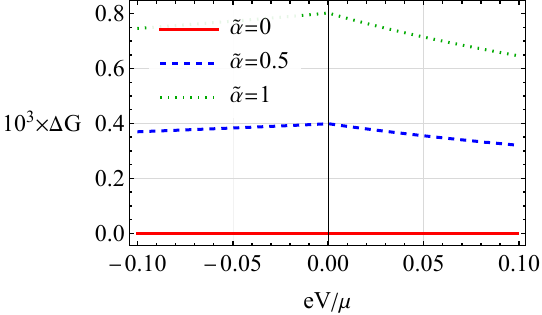}
\end{subfigure}
\caption{
The electric $G_e$ (panel (a)) and spin $G_{\sigma}$ (panel (b)) conductances as well as the conductance asymmetry $\Delta G$ (panel (c)). 
The spin-splitting vector $\tilde{\bm{\alpha}}$ is perpendicular to the interface, $\tilde{\alpha}_y=0$.
We use Eqs.~(\ref{8-i-G-e}), (\ref{8-i-G-s}), and (\ref{8-i-Delta-G-def}). In all panels, we set $\tilde{J}=1$ and $Z=10$.
}
\label{fig:8-i-J=1.-alphay=0-Z}
\end{figure*}

The barrier strength $Z$ suppresses the conductance and leads to a higher sensitivity of $G_e$ to the spin-splitting vector; see the corresponding conductances in Fig.~\ref{fig:8-i-J=1.-alphay=0-Z} for a relatively high $Z=10$. 

\begin{figure*}[!ht]
\centering
\begin{subfigure}[b]{0.31\textwidth}
         \includegraphics[width=\textwidth]{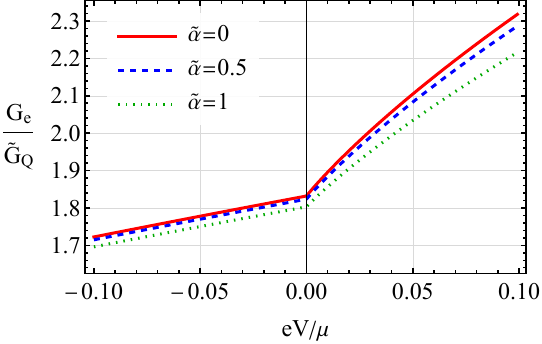}
         \caption{$\tilde{J}=1$}
\end{subfigure}
\begin{subfigure}[b]{0.31\textwidth}
         \includegraphics[width=\textwidth]{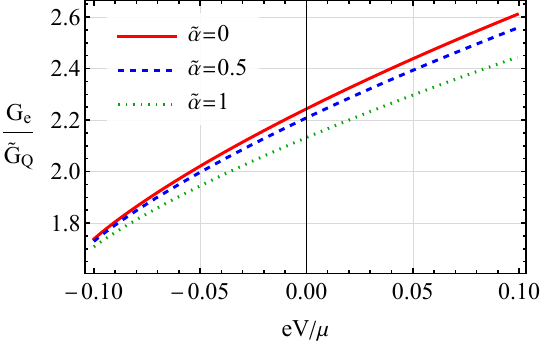}
         \caption{$\tilde{J}=0.9$}
\end{subfigure}
\caption{
The electric $G_e$ conductance for $\tilde{J}=1$ (panel (a)) and $\tilde{J}=0.9$ (panel (b)). The spin-splitting vector $\tilde{\bm{\alpha}}$ is parallel to the interface, $\tilde{\alpha}_x=0$. We use Eq.~(\ref{8-i-G-e}) and set $Z=0$.
}
\label{fig:8-i-alphax=0}
\end{figure*}

The results for the spin-splitting vector parallel to the interface, i.e., $\tilde{\alpha}_x=0$, are shown in Fig.~\ref{fig:8-i-alphax=0}. The spin conductance vanishes, and there is no spin-filtering in this case. Unlike the case $\tilde{\alpha}_y=0$, the electric conductance is more sensitive to the spin-splitting vector, cf. Figs.~\ref{fig:8-i-J=1-alphay=0} and \ref{fig:8-i-alphax=0}.

The absence of spin-filtering when the spin-splitting vector is parallel to the interface follows from a symmetry argument: the planar interface does not break the symmetry of the Fermi surface and the spin polarization $\varepsilon_{k_y,s} \leftrightarrow \varepsilon_{-k_y,-s}$. Then, the probability of tunneling from the metal into the parts of the Fermi surface in the $p$-wave magnet with a different spin polarization is the same, and, as a result, the conductance is insensitive to the spin of impinging particles. On the other hand, $\varepsilon_{k_x,s} \leftrightarrow \varepsilon_{-k_x,-s}$ is no longer a symmetry of the system. Therefore, we may expect different probabilities of tunneling for spin-up and spin-down electrons, as is confirmed by the results in Figs.~\ref{fig:8-i-J=1-alphay=0} and \ref{fig:8-i-J=1.-alphay=0-Z}.

\subsection{Two spin-polarized Fermi surface}
\label{sec:8-ii}

Let us contrast the spin-filtering in our model of unconventional $p$-wave magnets and in the model with two spin-polarized Fermi surfaces, which is reminiscent of a 1D spin-orbit coupled wire. We use the following Hamiltonian:
\begin{equation}
\label{8-ii-H-s}
H_{s}(x,k_y) =\left(-\frac{\nabla^2_{x}}{2m} +\frac{k_{\parallel}^2}{2m} -\mu\right)  + s\left[-\frac{i}{2}\left\{\alpha_x(x),\nabla_x\right\} +\alpha_y(x) k_{y}\right] +U(x),
\end{equation}
where $s=\pm$ is the spin projection, cf. Eq.~(\ref{8-i-H-lin}).
The Hamiltonian (\ref{8-ii-H-s}) corresponds to the antisymmetric spin-polarized spectrum shown in, e.g., Refs.~\cite{hayami2019momentum, Hayami-Kusunose-BottomupDesignSpinsplit-2020}. As we will demonstrate, this class of models has a different spin transport without the spin filtering effect. This observation may be useful in distinguishing different classes of magnets with antisymmetric spin splitting.

The scattered wave functions in the model (\ref{8-ii-H-s}) have the standard form
\begin{eqnarray}
\label{8-ii-psi-1}
\psi_s(x<0) &=& e^{ik_xx +ik_yy}+r_s e^{-ik_xx +ik_yy},\\
\label{8-ii-psi-2}
\psi_s(x>0) &=& t_s e^{iq_{x,s}x +ik_yy}
\end{eqnarray}
with $q_{x,s}=k_F \tilde{q}_{x,s}$ and
\begin{equation}
\label{8-ii-qx}
\tilde{q}_{x,s} = -\frac{s\tilde{\alpha}_x}{2}  +\sqrt{\tilde{\varepsilon} -s\tilde{\alpha}_y \tilde{k}_y +\frac{\tilde{\alpha}_x^2}{4} -\tilde{k}_y^2}.
\end{equation}
The boundary conditions are equivalent to those in Eqs.~(\ref{8-i-BC-1}) and (\ref{8-i-BC-2}) with $\sigma_z \to s$. 

By using the boundary conditions with the wave functions (\ref{8-ii-psi-1}) and (\ref{8-ii-psi-2}), we obtain the following reflection and transmission coefficients:
\begin{eqnarray}
\label{8-ii-r}
r_s &=& \frac{\tilde{k}_x -\tilde{q}_{x,s} -iZ -\frac{s\tilde{\alpha}_x}{2}}{\tilde{k}_x +\tilde{q}_{x,s} +iZ +\frac{s\tilde{\alpha}_x}{2}},\\
\label{8-ii-t}
t_s &=& \frac{2\tilde{k}_x}{\tilde{k}_x +\tilde{q}_{x,s} +iZ +\frac{s\tilde{\alpha}_x}{2}}.
\end{eqnarray}

The differential conductance at vanishing temperature is
\begin{equation}
\label{8-ii-G-def}
G_s(V) = L \int \frac{dk_y}{(2\pi)^2} \int d\varepsilon \left(1-|r_s|^2\right) \df{\varepsilon -\mu -eV} = \tilde{G}_{Q} \int d\tilde{k}_y \frac{2\left(\tilde{q}_{x,s} +s\tilde{\alpha}_x/2\right) \tilde{k}_x}{\left(\tilde{q}_{x,s} +\tilde{k}_x+s\tilde{\alpha}_x/2\right)^2 +Z^2} \Big|_{\tilde{\varepsilon} = eV/\mu +1},
\end{equation}
In calculating the conductance (\ref{8-ii-G-def}), the integration over $\tilde{k}_y$ is performed over a finite range determined from $\tilde{k}_x$ and $\tilde{q}_{x,s}$ being real; this corresponds to propagating waves.
Since the combination $\tilde{q}_{x,s}+s\tilde{\alpha}_x/2$ has the symmetry $s\to -s$ and $k_y \to -k_y$, the conductance is the same for both spin projections. Therefore, there is no spin filtering effect.

The absence of spin splitting can be explained by the symmetry of the Fermi surfaces in the metal and the magnet. The interface does not introduce any asymmetry between the overlap of spin-degenerate Fermi surface in the metal and the independent spin-split Fermi surfaces in the magnet.

\section{Nonlinear transport}
In this section, we calculate the electric and spin transport properties of bulk $p$-wave magnets. We use the semiclassical approach where the distribution function satisfies the following kinetic equation:
\begin{equation}
\label{8-kin-eq}
-e\mathbf{E}\cdot\partial_{\mathbf{k}} f = -\frac{f-f_{eq}}{\tau}.
\end{equation}
Here, $\mathbf{E}$ is the electric field, $f_{eq}=1/\left[e^{(\varepsilon_{\mathbf{k}} -\mu)/T}+1\right]$ is the equilibrium distribution function, $\varepsilon_{\mathbf{k}}$ is the dispersion relation, and $\tau$ is the relaxation time. In writing Eq.~(\ref{8-kin-eq}), we assumed a uniform and time-independent electric field. For simplicity, we focus on the case with a single filled band without any inter-sectoral coupling; in the effective model given in Eq.~(\ref{2-tmr-H-i-lin}), we set $|\mu|<|J|$. This allows us to neglect the inter-band scattering and use the simplest form of the collision integral for inelastic scattering in the relaxation-time approximation with a constant scattering time. In addition, since we can treat each of the sectors independently, we omit the sectoral index $\eta$ and only perform the summation over sectors in the final expressions. The latter summation also allows us to avoid calculating the spin-torque part of the spin current~\cite{Shi-Niu-ProperDefinitionSpin-2006, Tokatly-Tokatly-EquilibriumSpinCurrents-2008} which has the opposite sign in two sectors and, hence, cancels in the final result.

We solve Eq.~(\ref{8-kin-eq}) perturbatively in electric field:
\begin{equation}
\label{8-kin-f}
f = f_{eq} +f^{(1)} +f^{(2)} +\ldots,
\end{equation}
where
\begin{eqnarray}
\label{8-kin-f-1}
f^{(1)} &=& e \tau \left(\mathbf{E}\cdot\partial_{\mathbf{k}}\right)f_{eq},\\
\label{8-kin-f-2}
f^{(2)} &=& e \tau \left(\mathbf{E}\cdot\partial_{\mathbf{k}}\right)f^{(1)} = e^2 \tau^2 \left(\mathbf{E}\cdot\partial_{\mathbf{k}}\right)\left(\mathbf{E}\cdot\partial_{\mathbf{k}}\right)f_{eq}.
\end{eqnarray}
The perturbative approach is valid at $eE \tau v_F/\mu\ll1$, which is easily fulfilled in regular metals.

\subsection{First-order response}
\label{sec:8-kin-first}

The first-order contribution to the electric current density is defined as
\begin{equation}
\label{8-kin-j-1}
\mathbf{j}_{el}^{(1)} = -e \int \frac{d\mathbf{k}}{(2\pi)^2} \mathbf{v} f^{(1)}
 = -e^2 \tau \int \frac{d\mathbf{k}}{(2\pi)^2} \mathbf{v} \left(\mathbf{E}\cdot\partial_{\mathbf{k}}\right)f_{eq},
\end{equation}
where $\mathbf{v} = \partial_{\mathbf{k}} \varepsilon_{\mathbf{k}}$ is the group velocity.
We consider zero-temperature limit $T\to0$ and replace $\partial_{\mathbf{k}} \to \mathbf{v} \partial_{\varepsilon}$. Then, 
\begin{equation}
\label{8-kin-j-1-a}
\mathbf{j}_{el}^{(1)} = e^2 \tau \int \frac{d\mathbf{k}}{(2\pi)^2} \mathbf{v} \left(\mathbf{E}\cdot\mathbf{v}\right) \df{\varepsilon_{\mathbf{k}} -\mu}.
\end{equation}
In the case of the spin current density, we take into account the spin texture of the Fermi surface and replace the group velocity with the mean value of the spin current operator:
\begin{equation}
\label{8-kin-j-s-mean}
\left\langle \hat{\mathbf{j}}_{\sigma} \right\rangle = \psi^{\dag}\hat{\mathbf{j}}_{\sigma}\psi,
\end{equation}
where $\psi\equiv \psi_{\eta}$ is the eigenfunction of the effective Hamiltonian, see Eq.~(\ref{8-i-psi}). Then,
\begin{equation}
\label{8-kin-j-1-sigma}
\mathbf{j}_{\sigma}^{(1)} = e^2 \tau \int \frac{d\mathbf{k}}{(2\pi)^2} \left\langle \hat{\mathbf{j}}_{\sigma} \right\rangle \left(\mathbf{E}\cdot\mathbf{v}\right) \df{\varepsilon_{\mathbf{k}} -\mu}.
\end{equation}

For our model with the low-energy Hamiltonian (\ref{2-tmr-H-i-lin}), the energy spectrum is
\begin{equation}
\label{8-kin-i-eps}
\tilde{\varepsilon}_{\tilde{\mathbf{k}},\pm} = \tilde{k}^{2} \pm \sqrt{\tilde{J}^2 +\left(\tilde{\bm{\alpha}}\cdot\tilde{\bm{k}}\right)^2},
\end{equation}
where we used dimensionless variables. The spectrum is doubly degenerate in the sectoral degree of freedom $\eta$. We assume $\tilde{J}>1$: in this case, only the lower band $\tilde{\varepsilon}_{\tilde{\mathbf{k}},-}$ is filled, the upper band $\tilde{\varepsilon}_{\tilde{\mathbf{k}},+}$ is empty and can be disregarded in the DC transport. 

The group velocity for the lower band reads
\begin{eqnarray}
\label{8-kin-i-v}
\tilde{\mathbf{v}} = \left(\tilde{\mathbf{k}} +\frac{\tilde{\bm{\alpha}}}{2} S_{z} 
\right).
\end{eqnarray}
Here, $\tilde{v} =m v/k_F$ and
\begin{equation}
\label{8-kin-Sz-def}
S_{z} = \psi^{\dag} \sigma_z \psi =- \frac{(\tilde{\bm{\alpha}}\cdot\tilde{\mathbf{k}})}{\sqrt{\tilde{J}^2 + (\tilde{\bm{\alpha}}\cdot\tilde{\mathbf{k}})^2}}
\end{equation}
is the spin polarization per sector. In calculating the spin polarization (\ref{8-kin-Sz-def}), we used the wave functions $\psi$ defined in Eq.~(\ref{8-i-psi}) with $\tilde{\varepsilon} \to \tilde{\varepsilon}_{\tilde{\mathbf{k}},-}$; in addition, we omitted the subscript $\eta$. 

To determine the mean value of the spin current (\ref{8-kin-j-s-mean}), we use the eigenfunction $\psi$ in Eq.~(\ref{8-i-psi}) and the following spin current operator:
\begin{equation}
\label{8-kin-Sz-continuity}
\hat{\mathbf{j}}_{\sigma} = \frac{\mathbf{k}}{m} \sigma_z +\bm{\alpha}.
\end{equation}
The result for the lower band reads
\begin{equation}
\label{8-kin-i-j-s}
\left\langle \hat{\mathbf{j}}_{\sigma} \right\rangle  = \frac{k_F}{m} \left(\tilde{\mathbf{k}} S_{z} +\frac{\tilde{\bm{\alpha}}}{2}\right).
\end{equation}
Note that while the group velocity (\ref{8-kin-i-v}) is odd in momentum, $\left\langle \hat{\mathbf{j}}_{\sigma} \right\rangle$ is an even function. As we demonstrate below, this leads to a drastic difference between electric and spin current responses.

As follows from Eqs.~(\ref{8-kin-j-1-a}) and (\ref{8-kin-j-1-sigma}), where Eqs.~(\ref{8-kin-i-v}) and (\ref{8-kin-i-j-s}) were used, the electric and spin conductivities are
\begin{eqnarray}
\label{8-kin-i-2-sigma-ij-el}
\sigma_{ij}^{(el)} &=& 2\sigma_0 \sum_{\eta=\pm}\int_0^{\infty} \tilde{k} d\tilde{k} \int_0^{2\pi} d\theta 
\left(\tilde{k}_i +\frac{\tilde{\alpha}_i}{2}S_z\right)
\left[\tilde{k}_j -\frac{\tilde{\alpha}_j}{2} \frac{\left(\tilde{\bm{\alpha}}\cdot\tilde{\bm{k}}\right)}{\sqrt{\tilde{J}^2 +\left(\tilde{\bm{\alpha}}\cdot\tilde{\bm{k}}\right)^2}}\right]
\nonumber\\
&\times& \frac{\df{(\tilde{k} -\tilde{k}_{+})(\tilde{k} -\tilde{k}_{-})}}{\left|\partial_{\tilde{k}} \tilde{\varepsilon}_{-}\right|}
\end{eqnarray}
and
\begin{eqnarray}
\label{8-kin-i-2-sigma-ij-spin}
\sigma_{ij}^{(\sigma)} &=& 2\sigma_0 \sum_{\eta=\pm}\int_0^{\infty} \tilde{k} d\tilde{k} \int_0^{2\pi} d\theta 
\left(\tilde{k}_iS_z +\frac{\tilde{\alpha}_i}{2}\right)
\left[\tilde{k}_j -\frac{\tilde{\alpha}_j}{2} \frac{\left(\tilde{\bm{\alpha}}\cdot\tilde{\bm{k}}\right)}{\sqrt{\tilde{J}^2 +\left(\tilde{\bm{\alpha}}\cdot\tilde{\bm{k}}\right)^2}}\right] \nonumber\\
&\times& \frac{\df{(\tilde{k} -\tilde{k}_{+})(\tilde{k} -\tilde{k}_{-})}}{\left|\partial_{\tilde{k}} \tilde{\varepsilon}_{-}\right|},
\end{eqnarray}
respectively. Here, $\sigma_0 = e^2\tau \mu/\pi$ is the conductivity of a metal (i.e., at $\tilde{J}=0$ and $\tilde{\alpha}=0$).

For the sake of definiteness, we direct the $x$-axis along the spin-splitting vector $\tilde{\bm{\alpha}}$, $\left(\tilde{\bm{\alpha}}\cdot\tilde{\bm{k}}\right) = \tilde{\alpha} \tilde{k}\cos{\theta}$. Then,
\begin{equation}
\label{8-kin-k-pm-def}
\tilde{k}_{\pm} = \sqrt{1 +\frac{\tilde{\alpha}^2 \cos^2{\theta}}{2} \pm \sqrt{\tilde{J}^2-1 +\left(1 +\frac{\tilde{\alpha}^2 \cos^2{\theta}}{2}\right)^2 }}.
\end{equation}

Assuming $\tilde{\alpha}\ll1$, integrating over $\tilde{k}$, and expanding up to the second order in $\tilde{\alpha}$, we obtain
\begin{equation}
\label{8-kin-i-2-sigma-xx-el-app}
\sigma_{xx}^{(el)} \approx \sigma_0 \sum_{\eta=\pm}\int_0^{2\pi} d\theta 
\left(1+|\tilde{J}|\right) \cos^2{\theta} \left(1- \tilde{\alpha}^2\frac{5 - 3\cos{(2\theta)}}{8|\tilde{J}|}\right)
=\sigma_0  \left(1+|\tilde{J}| \right) \left(1 - \frac{7 \tilde{\alpha}^2}{16|\tilde{J}|}\right)
\end{equation}
and
\begin{equation}
\label{8-kin-i-2-sigma-yy-el-app}
\sigma_{yy}^{(el)} \approx \sigma_0 \sum_{\eta=\pm}\int_0^{2\pi} d\theta 
\left(1+|\tilde{J}|\right) \sin^2{\theta} \left(1+ \tilde{\alpha}^2\frac{3\cos^2{\theta}}{4|\tilde{J}|}\right)
=\sigma_0  \left(1+|\tilde{J}| \right) \left(1 +\frac{3\tilde{\alpha}^2}{16|\tilde{J}|}\right).
\end{equation}
The Hall conductivity $\sigma_{xy}^{(el)}$ and the spin conductivity tensor $\sigma_{ij}^{(\sigma)}$ vanish after the angular integration.

\begin{figure*}[!ht]
\centering
\begin{subfigure}[b]{0.45\textwidth}
         \includegraphics[width=\textwidth]{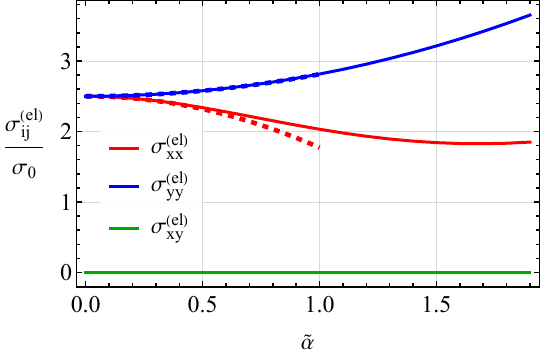}
\end{subfigure}
\caption{
The electric conductivity tensor as a function of the spin-splitting vector $\tilde{\alpha}$. We use Eq.~(\ref{8-kin-i-2-sigma-ij-el}) for exact results (solid lines) and Eqs.~(\ref{8-kin-i-2-sigma-xx-el-app}) and (\ref{8-kin-i-2-sigma-yy-el-app}) for the approximate results (dashed lines). The spin conductivity tensor vanishes $\sigma_{ij}^{(\sigma)}=0$. We set $\tilde{J}=1.5$.
}
\label{fig:8-kin-i-J=1.5}
\end{figure*}

We show the conductivity (\ref{8-kin-i-2-sigma-ij-el}) together with the asymptotes (\ref{8-kin-i-2-sigma-xx-el-app}) and (\ref{8-kin-i-2-sigma-yy-el-app}) in Fig.~\ref{fig:8-kin-i-J=1.5}. The decrease of the conductivity along the direction of $\tilde{\bm{\alpha}}$ can be understood from the band structure: with the rise of $\tilde{\alpha}$, the energy dispersion becomes more shallow along $\tilde{\bm{\alpha}}$ leading to the decrease of the group velocity; the trend is opposite in the direction perpendicular to $\tilde{\bm{\alpha}}$.

\subsection{Second-order response}
\label{sec:8-kin-second}

The second-order contribution to the electric current density is defined as
\begin{equation}
\label{8-kin-j-2}
\mathbf{j}^{(2)} = -e \int \frac{d\mathbf{k}}{(2\pi)^2} \mathbf{v} 
f^{(2)} = -e^3 \tau^2 \int \frac{d\mathbf{k}}{(2\pi)^2} \mathbf{v} \left(\mathbf{E}\cdot\partial_{\mathbf{k}}\right)\left(\mathbf{E}\cdot\partial_{\mathbf{k}}\right)f_{eq}.
\end{equation}
We find it convenient to separate two contributions in Eq.~(\ref{8-kin-j-2}):
\begin{eqnarray}
\label{8-kin-j-2-1}
\mathbf{j}^{(2,1)} &=& e^3 \tau^2 \int \frac{d\mathbf{k}}{(2\pi)^2}  \mathbf{v} \df{\varepsilon_{\mathbf{k}} -\mu}\left(\mathbf{E}\cdot\partial_{\mathbf{k}}\right) \left(\mathbf{E}\cdot\mathbf{v}\right),\\
\label{8-kin-j-2-2}
\mathbf{j}^{(2,2)} &=& e^3 \tau^2 \int \frac{d\mathbf{k}}{(2\pi)^2} \mathbf{v} \left(\mathbf{E}\cdot\mathbf{v}\right) \left(\mathbf{E}\cdot\partial_{\mathbf{k}}\right) \df{\varepsilon_{\mathbf{k}} -\mu}.
\end{eqnarray}

The derivative from the $\delta$-function can be tackled by using the following formula:
\begin{eqnarray}
\label{8-kin-delta-rel}
&&\int d\mathbf{k} F(\mathbf{k}) \partial_{k_i}\df{\varepsilon_{\mathbf{k}}-\mu} 
= \int d\varepsilon \int d\theta J(\varepsilon,\theta) v_i F(\varepsilon,\theta) \partial_{\varepsilon}\df{\varepsilon-\mu} \nonumber\\
&&= -\int d\varepsilon \int d\theta \df{\varepsilon-\mu} \partial_{\varepsilon}\left[J(\varepsilon,\theta) v_i F(\varepsilon,\theta)\right]
= -\int d\mathbf{k} \df{\varepsilon-\mu} \frac{1}{J(\varepsilon,\theta)} \partial_{\varepsilon}\left[J(\varepsilon,\theta) v_i F(\varepsilon,\theta)\right],
\end{eqnarray}
where $F(\mathbf{k})$ is a differentiable function of momentum, $J(\varepsilon,\theta) \equiv J(\mathbf{k}) = \tilde{k} \left|\frac{\partial \tilde{k}}{\partial \tilde{\varepsilon}}\right|$ is the Jacobian, and we integrated by parts in the second line.

Therefore, we have
\begin{equation}
\label{8-kin-j-2-2-a}
\mathbf{j}^{(2,2)} = -\frac{e^3 \tau^2}{(2\pi)^2} \int d\mathbf{k} \df{\varepsilon_{\mathbf{k}} -\mu} \frac{1}{J(\mathbf{k})} \partial_{\varepsilon}\left[J(\mathbf{k}) 
 \mathbf{v} \left(\mathbf{E}\cdot\mathbf{v}\right)  \left(\mathbf{E}\cdot\mathbf{v}\right) \right].
\end{equation}

For the spin current, we replace the first group velocity $\mathbf{v}$ in Eqs.~(\ref{8-kin-j-2-1}) and (\ref{8-kin-j-2-2}) with the mean value of the spin current, see Eq.~(\ref{8-kin-j-s-mean}).

The electric response tensor $\chi_{ijl}^{(el)}$ defined as $j_{el,i}^{(2)} = \chi_{ijl}^{(el)} E_jE_l$ is
\begin{equation}
\label{8-kin-i-chi-el-2}
\chi_{ijl}^{(el)} = \chi_0 \sum_{\eta=\pm}\int_0^{\infty} \tilde{k} d \tilde{k} \int_0^{2\pi} \frac{d\theta}{2\pi} \df{\tilde{\varepsilon} -1} 
\left[\tilde{v}_i \partial_{\tilde{k}_j} \tilde{v}_l
-\frac{2}{J(\tilde{k},\theta)} \partial_{\tilde{\varepsilon}}\left\{J(\tilde{k},\theta) \tilde{v}_i \tilde{v}_j \tilde{v}_l \right\}
\right],
\end{equation}
where $\chi_0 = e^3 \tau^2 k_F/m$ and we also included the summation over the two sectors; it leads to an overall factor of $2$. As required by the time-reversal symmetry, the response tensor $\chi_{ijl}^{(el)}$ vanishes.

For the spin response tensor $\chi_{ijl}^{(\sigma)}$ defined as $j_{\sigma,i}^{(2)} = \chi_{ijl}^{(\sigma)} E_jE_l$, components $\chi_{ijl}^{(\sigma,1)}$ and $\chi_{ijl}^{(\sigma,2)}$, we have
\begin{equation}
\label{8-kin-i-chi-spin-2}
\chi_{ijl}^{(\sigma)} = \chi_0 \sum_{\eta=\pm}\int_0^{\infty} \tilde{k} d \tilde{k} \int_0^{2\pi} \frac{d\theta}{2\pi} \df{\tilde{\varepsilon} -1} 
\left\{\left(\tilde{k}_i +\frac{\tilde{\alpha}_i}{2}S_z\right)  \partial_{\tilde{k}_j} \tilde{v}_l
-\frac{2}{J(\tilde{k},\theta)} \partial_{\tilde{\varepsilon}}\left[J(\tilde{k},\theta) \left(\tilde{k}_i +\frac{\tilde{\alpha}_i}{2}S_z\right)  \tilde{v}_j \tilde{v}_l \right]
\right\}.
\end{equation}
We use Eq.~(\ref{8-kin-i-chi-spin-2}) in Fig.~4 in the main text.

\bibliography{main}